%% file: paper.tex
\documentclass[journal=jacsat,manuscript=article]{achemso}

\usepackage{chemformula} 
\usepackage[T1]{fontenc} 
\usepackage{amsmath}
\usepackage{amssymb}
\usepackage{mathtools}
\usepackage{amsthm}
\usepackage{multirow}
\usepackage{amsthm}
\usepackage{microtype}
\usepackage{graphicx}
\usepackage{paralist, tabularx}
\usepackage{booktabs}
\usepackage{algorithm}
\usepackage{algpseudocode}
\usepackage{subfiles}       
\usepackage{stackrel}
\usepackage{caption}
\usepackage{subfig}

\usepackage{lmodern}

\theoremstyle{plain}

\theoremstyle{definition}

\theoremstyle{remark}

\author{Nhat Khang Ngo}
\affiliation{FPT Software AI Center, Hanoi, Vietnam}
\altaffiliation{Equal Contribution}
\author{Truong Son Hy}
\altaffiliation{Equal Contribution}
\affiliation{Department of Mathematics and Computer Science, Indiana State University, Terre Haute, IN 47809, USA}
\alsoaffiliation{Corresponding Author}
\email{TruongSon.Hy@indstate.edu}

\title[An \textsf{achemso} demo]
  {Target-aware Variational Auto-encoders for Ligand Generation with Multimodal Protein Representation Learning}

\abbreviations{IR,NMR,UV}

\begin{document}

\begin{tocentry}

Some journals require a graphical entry for the Table of Contents.
This should be laid out ``print ready'' so that the sizing of the
text is correct.

Inside the \texttt{tocentry} environment, the font used is Helvetica
8\,pt, as required by \emph{Journal of the American Chemical
Society}.

The surrounding frame is 9\,cm by 3.5\,cm, which is the maximum
permitted for  \emph{Journal of the American Chemical Society}
graphical table of content entries. The box will not resize if the
content is too big: instead it will overflow the edge of the box.

This box and the associated title will always be printed on a
separate page at the end of the document.

\end{tocentry}



\input{abstract}
\input{introduction}
\input{related}
\input{background}
\input{method}

\input{results}
\input{data_and_software}

\bibliography{paper}

\end{document}

%% file: abstract.tex
\begin{abstract}
Without knowledge of specific pockets, generating ligands based on the global structure of a protein target plays a crucial role in drug discovery as it helps reduce the search space for potential drug-like candidates in the pipeline. However, contemporary methods require optimizing tailored networks for each protein, which is arduous and costly. To address this issue, we introduce \textbf{TargetVAE}, a target-aware variational auto-encoder that generates ligands with high binding affinities to arbitrary protein targets, guided by a novel multimodal deep neural network built based on graph Transformers as the prior for the generative model. This is the first effort to unify different representations of proteins (e.g., sequence of amino-acids, 3D structure) into a single model that we name as \textbf{Protein Multimodal Network} (PMN). Our multimodal architecture learns from the entire protein structures and is able to capture their sequential, topological and geometrical information. We showcase the superiority of our approach by conducting extensive experiments and evaluations, including the assessment of generative model quality, ligand generation for unseen targets, docking score computation, and binding affinity prediction. Empirical results demonstrate the promising performance of our proposed approach. Our software package is publicly available at \url{https://github.com/HySonLab/Ligand_Generation}.
\end{abstract}

%% file: introduction.tex
\section{Introduction}

Drug discovery is a complex and expensive process that involves multiple stages and often takes years of development, with costs running into billions of dollars \cite{hughes2011principles}. The first stage is to design novel drug-like compounds that have high binding affinities to protein targets. This process consists of two sub-tasks: searching for candidates and measuring drug-target affinities (DTA). Searching for potential candidates in a huge database of roughly $10^{33}$ chemically valid molecules is a daunting task as current methods often rely on virtual screenings, professional software, and expert evaluation \cite{verkhivker2001binding, burley2019rcsb}. Besides, drug-target affinities (DTA) are critical measurements for identifying potential candidates, as well as avoiding those that are inefficient for clinical trials. The most reliable technique for predicting DTA involves atomistic molecular dynamics simulations. However, these methods are computationally expensive and time-consuming, making them infeasible for large-scale sets of protein-ligand complexes. Our ultimate objective is to accelerate and automate these two sub-tasks in the first stage of the drug development process, using computational methods and machine-learning techniques. 

To effectively design probable drug-like candidates, deep generative models  \cite{NEURIPS2018_d60678e8, pmlr-v80-jin18a, pmlr-v119-jin20a,luo2021a,Simonovsky2018GraphVAETG, de2018molgan, pmlr-v139-luo21a} have been proposed as a potential approach to reduce the amount of work for wet-lab experiments \cite{gapsys2022pre, burley2019rcsb, verkhivker2001binding}. These methods demonstrate remarkable results in  the unconditional generation or optimization for simple molecular properties (e.g., QED, SA, etc.). However, when enhancing binding affinity or other computationally expensive molecular properties, these generative models are prohibitively slow. They need to be trained in reinforcement learning frameworks where the generated molecular graph is modified based on the reward. It is worth noting that this reward function is determined by calling a property network that estimates the binding affinities. Albeit effective and powerful, these approaches require that specific property networks are trained for each protein target, which is not trivial due to the vast amount of (un)-known proteins \cite{berman2000protein, burley2019rcsb}. Furthermore, binding scores (labels) for supervision training are not widely available, and computing them via software like Autodock or Vina is time-consuming.

Proteins are macromolecules that can be represented in terms of sequences of amino acids (i.e. primary structure), 2D graphs at residue level constructed by nearest neighbors from folding information, or 3D point clouds at atom level. Recent advanced methods for protein representation learning leverage language models, graph neural networks (GNNs), and convolutional neural networks (CNNs). In sequence-based methods \cite{pmlr-v162-notin22a, 10.1093/bioinformatics/btac020, asgari2019probabilistic, WU202118, 10.1093/bioinformatics/bty178}, a protein sequence is regarded as a long sequence of tokens (i.e. $k$-mers \cite{pmlr-v162-notin22a}) that are fed to a transformer-based language model. In contrast, GNNs and CNNs-based approaches \cite{NEURIPS2019_03573b32, townshend2020atom3d, jing2021equivariant, jing2021learning, 10.1093/nargab/lqac004} operate on relational and geometric structures of proteins, respectively. While sequence-based approaches can capture the relationships among distant residues in a long protein sequence, they are not able to exploit the geometric relations among them. On the other hand, although GNNs and CNNs can learn spatial information about protein structures, they are limited in their ability to capture long-range interactions in large protein structures due to their reliance on localities.

\paragraph{Contributions}In summary, our contributions are three-fold as follows:
\begin{compactitem}
    \item We build a conditional VAE model, named \textbf{TargetVAE}, that can generate chemically valid, drug-like molecules with high binding scores to an arbitrarily given protein structure. Apart from other methods, ours can directly condition on the entire structure of any protein target and design multiple candidates that can bind to it, without requiring the training of a specific property network for each target. It is important to note that our generative model is conditional on the whole protein structure rather than a specific pocket or binding site (i.e. a region of the protein surface where the ligand binds to) as in \cite{guan3d, pmlr-v162-peng22b, NEURIPS2021_31445061, pmlr-v162-liu22m}. Our approach is more flexible because a protein complex can contain multiple binding sites and therefore potentially have different ligands. TargetVAE allows us to efficiently generate ligand candidates with high binding-affinity without the prior knowledge of any binding site.
    \item To diversify the generated results, we adapt previous works in computer vision domains. Specifically, we aim at transferring weights of a pre-trained unconditional VAE, which is trained on a large dataset of drug-like molecules, to a conditional VAE which is trained on a small, well-aligned dataset of protein-ligand pairs, allowing us to generate diverse sets of molecules while keeping relevant to the reference targets.
    \item We design a novel architecture, named \textbf{Protein Multimodal Network} (PMN), that unifies different modalities of proteins, i.e. sequences and 3D structures, to improve the performance of predicting binding affinities. The proposed model shows competitive results on two DTA benchmarks in comparison with state-of-the-art methods. Our novel multimodal architecture enables us to efficiently produce a protein embedding and accurately estimate protein-ligand binding affinity, and potentially replace the computationally expensive molecular dynamical simulation in the evaluation of ligand generation.
\end{compactitem}


%% file: related.tex
\section{Related Work}
\paragraph{Protein-Ligand Binding Prediction}
Machine learning methods, especially graph neural networks, have emerged as effective techniques for binding affinity prediction \cite{Scantlebury2023}. Using a Kronecker regularized least squares approach (KronRLS), \citet{nascimento2016multiple} cast the problem as a link prediction task on  bipartite networks and compute distinct kernels that indicate the similarities among drugs and targets to make predictions. Apart from binary prediction, \citet{He2017} introduce a framework named SimBoost that can predict continuous binding affinity scores between drugs and targets. In the deep learning era, \citet{10.1093/bioinformatics/bty593} propose DeepDTA, a deep-learning-based method that uses convolutional neural networks to operate on sequence representations of drugs and protein targets. Moreover, \citet{Zhao2020-gx} use generative adversarial networks (GANs) to learn better feature representations for compounds and proteins. On the other hand, \citet{10.1093/bioinformatics/btaa921}; \citet{D3RA00281K} utilize graph neural networks to learn the representations of the molecular graphs and protein structures, which are superior to the previous methods operating on sequences and handcrafted features. However, the common limitation of these above approaches is that none of them covers a wide range of representations of proteins. Indeed, our proposed method is the first multimodal model combining sequence, graph, and spatial information of proteins together.

\paragraph{Molecule Generation} Previous studies on molecule generation are mostly categorized into SMILES-based and graph-based approaches \cite{Merz2020}. \citet{doi:10.1021/acscentsci.7b00572, doi:10.1021/acscentsci.7b00512, Gao2020} use recurrent neural networks to build generative models operating on SMILES strings. However, these SMILES-based approaches often generate chemically invalid molecules. \citet{pmlr-v70-kusner17a} and \citet{dai2018syntaxdirected} circumvent this issue by augmenting the decoders with grammar and semantic constraints to only generate valid molecules, yet this added information does not fully capture chemical validity in the latent space. Apart from SMILES-based methods, several works \cite{Simonovsky2018GraphVAETG,pmlr-v119-jin20a, de2018molgan, pmlr-v80-jin18a, pmlr-v139-luo21a, luo2021a, thiede2020general, Hy_2023} propose graph generative models to design novel drug-like molecules. For example, \citet{pmlr-v80-jin18a}; \citet{pmlr-v119-jin20a} can generate molecules with 100$\%$ validity, but their methods are relatively slow as chemical rules are verified during the generative process. Although graph-based methods perform well in unconditional molecule generation, they have difficulty generating molecules with optimized properties. For a target-aware generation, reinforcement learning methods are used to systematically modify the generated molecular graphs \cite{NEURIPS2018_d60678e8, pmlr-v80-jin18a, pmlr-v119-jin20a, luo2021a}. Distinguishing from other prior works based on SMILES representation, we utilize the recently proposed SELFIES representation \cite{Krenn_2020} to achieve a high chemically validity in generated ligands. Given the success of conditional VAE in image generation \cite{harvey2022conditional}, our work is the first attempt to introduce a learnable prior based on the whole protein structures.

%% file: background.tex
\section{Background}

\subsection{Rotational Invariant Features}
\label{background:rotation_invariant}
According to \citet{jing2021learning}, geometric features of a residue node on the protein structure can be represented as a tuple $(s, V)$ of scalar features $s \in \mathbb{R}^n$ and vector features $V \in \mathbb{R}^{\mu \times 3}$. Respectively, $s$ and $V$ are invariant and equivariant with respect to geometric transformations in Euclidean space. In addition, to make the information propagate effectively from the vector channel to the scalar channel, \citet{jing2021learning} propose geometric vector perceptron (GVP) as a replacement for conventional dense layers in graph neural networks, enabling them to operate on geometric vectors and structural information of 3D large protein graphs.

The module transforms an input tuple $(s, V)$ of scalar features $s \in \mathbb{R} ^ {n}$ and vector features $V \in \mathbb{R} ^ {\mu \times 3}$ into a new tuple $(s^\prime, V^\prime) \in \mathbb{R}^m \times \mathbb{R}^{\nu \times 3}$. According to Algorithm \ref{algo:gvp}, GVP consists of two separate linear transformations $W_m$ and $W_h$ that work on the scalar and vector features respectively, followed by nonlinearities $\sigma$ and $\sigma^+$. Before being transformed, the scalar feature $s$ is concatenated with the $L_2-\text{norm}$ of the vector feature $V$. This enables GVP to extract the rotation-invariant information from the input vector $V$. Moreover, an additional transformation $W_{\mu}$ is used to control the dimensionality of the output vector $V^\prime$, making it independent of the number of norms extracted. Albeit simple, GVP is an effective module that guarantees desired properties of invariance/equivariance and expressiveness. The scalar and vector outputs of GVP are invariant and equivariant respectively, with respect to an arbitrary composition $R$ of rotations and reflections in 3D Euclidean space. In other words, if $\text{GVP}(s, V) = (s^\prime, V^\prime)$, then $\text{GVP}(s, R(V)) = (s^\prime, R(V^\prime))$.

\begin{algorithm}[h]

\caption{Geometric Vector Perceptron}\label{alg:cap}
\begin{algorithmic}
\State \textbf{Input:} Scalar and vector features $(s, V) \in \mathbb{R} ^ n \times \mathbb{R} ^ {\mu \times 3} $ 
\State \textbf{Output:} Scalar and vector features $(s^\prime, V^\prime) \in \mathbb{R} ^ m \times \mathbb{R} ^ {\nu \times 3} $ 
\State $h \gets \text{max}(\mu, \nu)$ 
\State \textbf{GVP}:
\indent
\State $V_h \gets W_h V \quad \in \mathbb{R} ^ {h \times 3}$
\State $V_\mu \gets W_\mu V_h \quad \in \mathbb{R} ^ {\mu \times 3}$
\State $s_h \gets \|V_h\|_2 \text{ (row-wise)} \quad \in \mathbb{R} ^ h$
\State $v_\mu \gets \|V_\mu\|_2 \text{ (row-wise} \quad \in \mathbb{R} ^ \mu$
\State $s_{h+n} \gets \text{concat}(s_h, s) \quad \in \mathbb{R} ^ {h + n}$
\State $s_m \gets W_m s_{h+n} + b \quad \in \mathbb{R} ^ m$
\State $s ^ \prime \gets \sigma(s_m) \quad \in \mathbb{R} ^ m$
\State $V^\prime \gets \sigma^+(v_\mu) \odot V_\mu \text{ (row-wise multiplication)} \quad \in \mathbb{R} ^ {\mu \times 3}$
\State \textbf{return} $(s^\prime, V ^ \prime)$
\end{algorithmic}
\label{algo:gvp}
\end{algorithm}

\subsection{Variational Auto-Encoders}
A variational auto-encoder (VAE) is regarded as an auto-encoding variational Bayes model \cite{kingma2013auto} that comprises two components, including a generative model and an inference model (also known as probabilistic encoder). The former uses a probabilistic decoder $p_\theta(x | z)$ and a prior $p_\psi(z)$ to define a joint distribution $p_{\theta, \psi}(x, z) = p_\theta(x | z)p_\psi(z)$ between latent variables $z$ and data $x$; in addition, \citet{kingma2013auto} let $p_\psi(z)$ be isotropic Gaussian. An ideal generative model should learn to maximize the log-likelihood $\log p_{\theta, \psi}(x) = \log \int p_{\theta, \psi}(x, z)dz$. However, this is intractable as marginalization over the latent space is usually infeasible with realistic data. VAE alleviates this issue by using an encoder $q_\phi(z |x)$ to approximate the true posterior distribution of the latent space and maximize the evidence lower bound (ELBO) over each training sample $x$: 
\begin{equation}
    \log p_{\theta, \psi}(x) \ge \mathbb{E}_{q_\phi}[\log p_{\theta, \psi}(x | z)] - \text{KL}[q_\phi(z | x) || p_\psi(z)].
\end{equation}

\vskip 0.25em
In conditional VAE, the generative component is augmented by auxiliary covariates $y$. Given a condition $y$, the generative model defines a conditional joint distribution of $z$ and $x$ as $p_{\theta, \psi}(x, z | y) = p_{\theta}(x | y,z)p_{\psi}(z | y)$. Similarly, the condition inputs are integrated into the encoder as $q_{\phi}(z|x,y)$. These two extensions establish a prominent conditional VAE model \cite{NIPS2015_8d55a249,Zheng_2019_CVPR,ivanov2018variational,wan2021high} that is trained to maximize the conditional ELBO as:
\begin{equation}
\label{conditional_elbo}
    \log p_{\theta, \psi}(x | y) \ge O_\text{cond} \triangleq  \mathbb{E}_{q_\phi}[\log p_{\theta, \psi}(x | y, z)] - \text{KL}[q_\phi(z | x,y) || p_\psi(z|y)].
\end{equation}







%% file: method.tex

 \section{Protein Multimodal Network (PMN)}
 \label{method:pmn}
 \begin{figure}[t] 
     \centering
     \includegraphics[scale = 0.3]{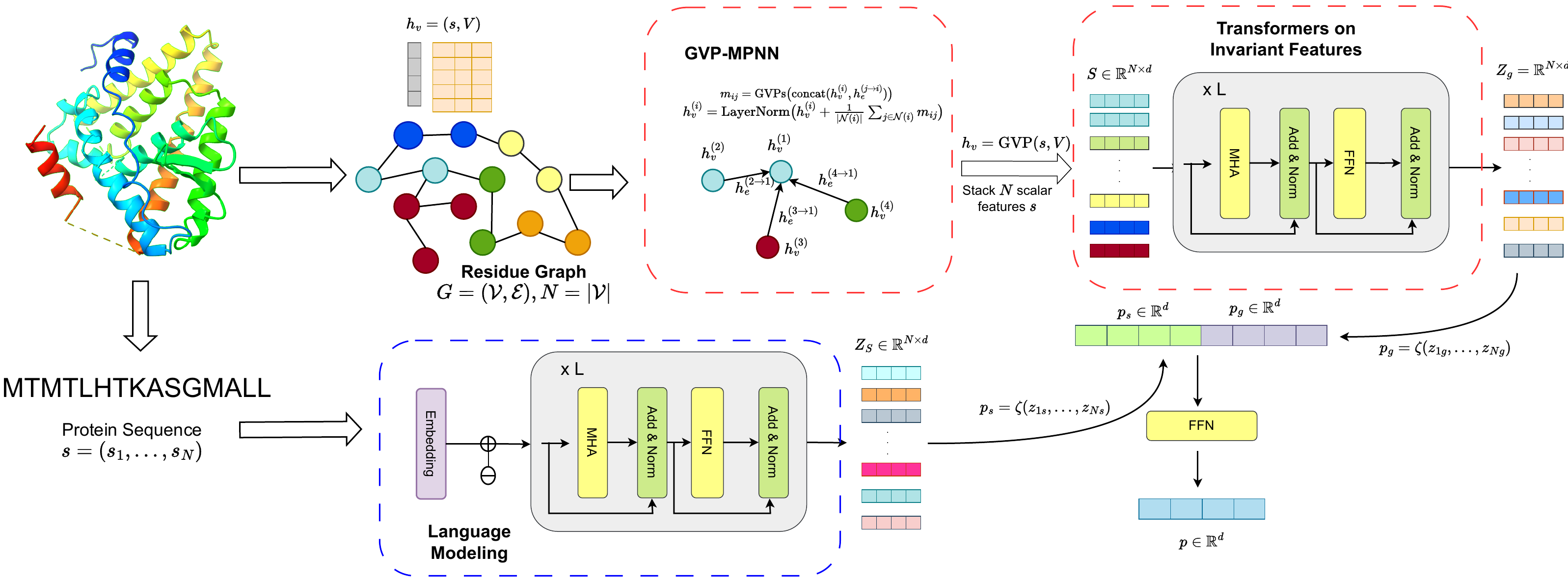}
     \caption{Overview of our Protein Multimodal Network (PMN).} 
     \label{fig:protein_model}
 \end{figure}


 Proteins are complex structures that consist of long chains of residues/amino acids. Each amino acid is a molecule with 3D structures, and a combination of hundreds to thousands of residues determines the unique 3D structure of a specific protein and its functions. It is worth noting that while two residues are distant along the protein sequence, they could be close to each other in three-dimensional space. This is our key observation to design a novel framework that can unify different representations of proteins in an end-to-end learning manner. In the field of graph learning, the conventional graph neural networks based on the message passing scheme \cite{10.5555/3305381.3305512} that propagates and aggregates information of each node to and from its local neighborhoods have been shown to be incapable of capturing the long-range interactions in a large-diameter or long-range \footnote{Diameter of a graph is defined as the maximum length of the shortest paths among all pairs of nodes.} graph \cite{NEURIPS2022_8c3c6668, 10.1063/5.0152833} while suffering from over-smoothing \cite{Chen_Lin_Li_Li_Zhou_Sun_2020} and over-squashing \cite{topping2022understanding} problems. Furthermore, to obtain the global understanding of the whole input graph, the message passing scheme needs the number of layers / iterations proportional to the diameter length for distant nodes to ``communicate'' with each other. That is computationally infeasible for macromolecules with thousands of atoms or residues like proteins. Meanwhile, the graph Transformers that considers all pairwise node interactions via the self-attention mechanism can successfully capture the long-range dependencies \cite{kim2022pure, cai2023connection, 10.1063/5.0152833}. Since proteins can be seen as long-range graphs, we utilize sequential and graph Transformers to encode both sequences and 3D graphs of residues and combine them to create a unified representation for a large protein, making our model operate on multi-modalities of proteins. Our Transformer-based model can efficiently capture both the local and global information of a protein with a reasonably small number of layers. In this section, we regard $W_*$ and $b_*$ as learnable weight matrix and bias vector of a linear layer, respectively.

 \paragraph{Long-range Modeling on 3D Structures} 
According to Figure \ref{fig:protein_model}, there are three major components in the 3D modeling part, including a local encoder, a GVP module, and a global Transformers encoder (Trans). We use a message-passing network (MPNN) in which dense layers are replaced by GVP to operate on invariant features \cite{jing2021learning}:
\begin{align}
   m_{ij} & = \text{GVPs}\big(\text{concat}(h ^ {(i)} _v, h ^ {(j \rightarrow i)} _e) \big), \\
    h ^ {(i)} _v & = \text{LayerNorm}\bigg(h ^ {(i)} _v + \frac{1}{|\mathcal{N}(i)|}\sum_{j \in \mathcal{N}(i)} m_{ij}\bigg).
\end{align}
Where $m_{ij}$ computed by a module of three GVP layers denotes the message propagated from node $j$ to $i$. Also, $h ^ {(i)} _v$ and $h ^ {(j \rightarrow i)} _e$ indicate the embeddings of node $i$ and edge $(j \rightarrow i)$ and are tuples of scalar and vector features as described in Section \ref{background:rotation_invariant}. The local encoder outputs a tuple of scalar and vector features for each residue node, which are rotationally invariant and equivariant, respectively. We utilize a GVP module to update the tuple $h_v = (s, V)$ as $(s',V') = \text{GVP}\big((s, V)\big)$, and we take the invariant scalar feature $s' \in \mathbb{R} ^ d$ as the node embedding for successor modules. The resulting tensor $S \in \mathbb{R}^{N \times d}$, in which row $i$ indicates a $d$-dimensional scalar feature $s_i$ of node $i$, is passed to a $L$-layer Transformers encoder:
\begin{align}
    Q_\ell &= Z_{\ell-1} W_\ell ^ Q,  \ \ K_\ell = Z_{\ell-1} W_\ell ^ K, \ \ V_\ell = Z_{\ell-1} W_\ell ^ V ,\label{equation_5}\\
    H_\ell &= \text{MultiheadAttention}(Q_\ell, K_\ell, V_\ell) ,    \label{equation_6}\\
    Z_\ell &= \text{LayerNorm}(Z_{\ell-1} + \text{FFN}(H_\ell)).
    \label{equation_7}
\end{align}
Here, we initialize $Z_0 \triangleq S$; and we have $\{W_\ell ^ Q, W_\ell ^ K, W_\ell ^ V\}_{\ell=1} ^ L \in \mathbb{R} ^ {d \times d_k}$ as learnable weight matrices / parameters corresponding to the query $Q_\ell$, key $K_\ell$ and value $V_\ell$ matrices, respectively, of the Transformer at each layer $\ell$; and $Z_g \triangleq Z_L$ denotes the final node embeddings produced by the network. 
Notably, this global encoder allows residue nodes to attend to other nodes on a large protein graph, especially those that are distant from them (i.e. long-range modeling). Finally, we aggregate node embeddings by a row-wise \textit{Aggregator} $\zeta$ (e.g., mean, max, sum, etc.) to produce an embedding for the protein structure $p_g = \zeta(Z_g) \in \mathbb{R} ^  d$.
 
\paragraph{Language Modeling on Protein Sequence}
A protein can be represented as a sequence $s = (s_0, s_1, ..., s_n)$ in which $s_i \in \mathbb{R}^{20}$ is a one-hot vector indicating one in a total of 20 types of residues. We utilize Transformer-based language models, where the layers are the same as in Eq. (\ref{equation_5}, \ref{equation_6}, \ref{equation_7}), to compute the text representation of this protein sequence with the initial embeddings $Z_0 = [z_1,z_2, ..., z_n] \in \mathbb{R} ^ {n \times d}$ with $z_i \in \mathbb{R}^d$ is calculated as $z_i = \text{Embed}(s_i) + p_i$. 

Here, $p_i$ is the positional encoding feature added at each token $i$. Then, we define $p_s = \zeta(Z_s) \in \mathbb{R} ^ d$ as the global representation for the entire protein sequence. Notably, there may be hundreds to thousands of residues in a long-chain protein, so we utilize efficient Transformers \cite{roy*2020efficient, choromanski2021rethinking, Kitaev2020Reformer:} to reduce the computational complexity. At the end of the network, we calculate a unified representation of the protein $p = W_2 \text{ReLU}(W_1 (\text{concat}(p_g,p_s)) + b_1) + b_2$.

\section{Binding Affinity Prediction and Ligand Generation}
\subsection{Problem Setup}
\begin{figure}
    \centering
    \includegraphics[scale=0.75]{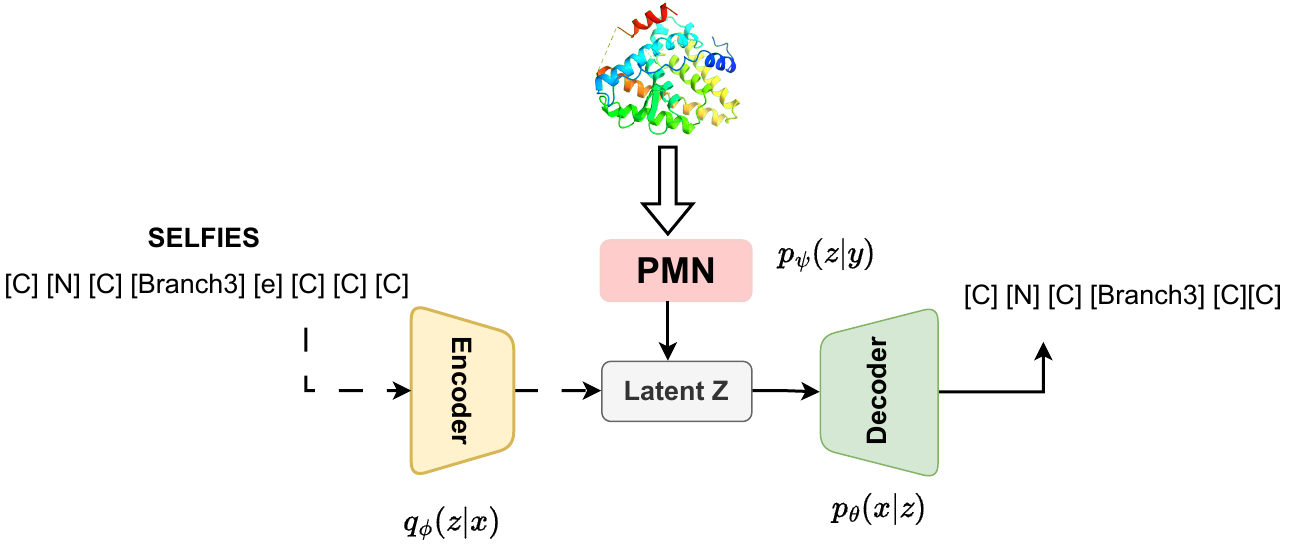}
    \caption{TargetVAE with an encoder, decoder, and a prior network. The PMN prior network computes the conditions from protein structures for constructing the latent space of the VAE framework, which learns to generate SELFIES representations of molecules.}
    \label{fig:ligand_generation}
\end{figure}

Given a well-aligned dataset $D$ of protein-ligand pairs, our objective is to predict the binding affinity  and generate novel drug-like ligands that have the potential to bind to a conditioning protein structure. We cast the former as a prediction task based on geometric and relational reasoning on protein and ligand structures, whereas the latter is regarded as a protein-structure conditioned ligand generation. Let $(l,p,s) \in D$ be a pair of protein-ligand where $l$ and $p$ denote the representations of ligands and proteins, respectively, and $s$ indicates the binding score between them. We define such representations for proteins and ligands that best fit with the corresponding objectives in the following sections. Additionally, Figures \ref{fig:binding_prediction} and \ref{fig:ligand_generation} depict the overview of our approach in both tasks.
\subsection{Binding Affinity Prediction}


\begin{figure}
    \centering
    \includegraphics[scale=0.85]{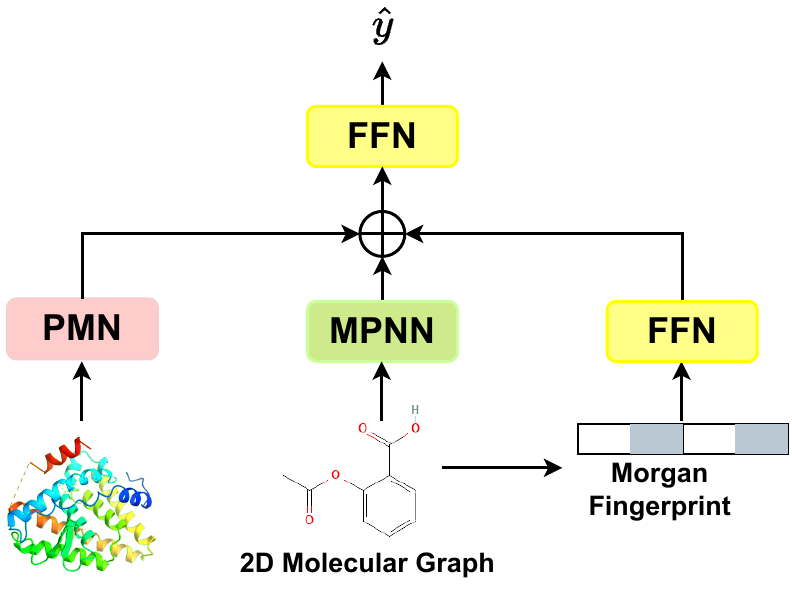}
    \caption{A framework for predicting binding affinities between proteins and ligands.}
    \label{fig:binding_prediction}
\end{figure}

\noindent 
Figure \ref{fig:binding_prediction} illustrates our proposed approach to predicting the binding affinities between ligands $l$ and protein targets $p$. A drug-like ligand is represented by a 2D molecular graph $G_l$ and a binary Morgan Fingerprint vector $v_{m} \in \mathbb{R}^{2,048}$\cite{rogers2010extended}, which embodies critical properties of chemical structures. $G$ and $v_{m}$ are passed to a message-passing neural network (MPNN) and feed-forward network (FFN). Also, the given protein structure $p$ is sent to the protein multimodal network (PMN):
\begin{align}
    z_{l1} &= \text{MPNN}(G_l),\\
    z_{l2} &= W_{m2} \text{ReLU}(W_{m1} v_m + b_{m1}) + b_{m2},\\
    z_{p} &= \text{PMN}(p). 
\end{align}
Here, $z_{l1}$ and $z_{l2}$ are the final embeddings of the ligand $l$, which are computed by a message-passing neural network (MPNN) from the ligand's 2D graph $G_l$ and a feed-forward neural network (FFN) from the global Morgan Fingerprint feature $v_m$, respectively. Then, $z_{l1}$, $z_{l2}$, and $z_{p}$ are combined to yield a unified input for the top FFN to output a scalar value $\hat{y}$ denoting the predicted binding affinity score:
\begin{align}
    \hat{y} &= W_{u2} \text{ReLU}(W_{u1} \text{concat}(z_{l1}, z_{l2}, z_{p}) + b_{u1}) + b_{u2}.
\end{align}

\subsection{Target-aware Ligand Generation}
Although there exist many machine-learning approaches that generate drug-like molecules, it is challenging for graph-based or smiles-based methods to generate chemically valid ligands with high probability. Meanwhile, SELFIES (SELF-referenced Embedded Strings) \cite{Krenn_2020} is a string-based representation of molecules that is 100$\%$ robust to molecular validity. A ligand $l$ can be defined as a string of $l = (l_1, l_2, ..., l_n)$ in which $l_i$ is a SELFIES token, which belongs to a predefined symbol set $S$ derived from the training dataset. We generate ligands  $\hat{l} = (\hat{l}_1, \hat{l}_2, ..., \hat{l}_n)$  by computing $n$ independent probability vectors $y = (y_1, y_2, ..., y_n)$, $y_i \in \mathbb{R}^{|S|}$. Each new token $\hat{l}_i$ is defined as $\hat{l}_i = S_j$ where $j=\stackrel[0 \le j < |S|]{}{\mathrm{argmax}} (y_i)$. 

Let $\phi, \theta$, and $\psi$ denote the encoder, decoder, and prior network in a conditional VAE framework, respectively. According to Figure \ref{fig:ligand_generation}, in this work, $\phi: \mathbb{R} ^ {n \times |S|} \rightarrow \mathbb{R} ^ {d}$ and $\theta:\mathbb{R}^{d} \rightarrow \mathbb{R} ^ {n \times |S|}$ are multiplayer perceptrons (MLPs), and $\psi$ is the PMN described in Section \ref{method:pmn} where the language modeling part is excluded for computational efficiency. All the networks are jointly optimized based on Equation \ref{conditional_elbo}. After training, given a protein structure $p$, a ligand $\hat{l}$ is generated by sampling a latent vector $z \sim \mathcal{N}(\mu_{\psi}(p), \sigma_{\psi}(p))$, which is fed to the decoder $\theta$ to decode into a SELFIES representation.
\vskip 0.25em
\paragraph{Conditional Inference with Pretrained Unconditional VAE}
\vskip 0.25em
In addition to validity, the diversity of generated sets of ligands is also an important criterion in drug discovery. While classical conditional VAE trained on protein-ligand pairs can generate novel and valid molecules, the diversity and uniqueness of these samples are relatively low due to the limited amount of available data. We address this issue by adapting the work proposed in \cite{harvey2022conditional} from the computer vision domain to diversify the latent variables. In this framework, the decoder $\theta$ of the generative model is independent with the condition $y$ as $p_{\theta, \psi}(x, z | y) = p_{\theta}(x | z) p_{\psi}(z | y)$, allowing $\theta$ to re-use weights of the decoder $\theta^*$ of an unconditional VAE as both have the identical architecture. We train the model to optimize the objective as:
\begin{equation}
\label{relax_conditional_elbo}
    \log p_{\theta, \psi}(x | y) \ge O_\text{for} \triangleq \mathbb{E}_{q_\phi}[\log p_{\theta, \psi}(x | z)] - \text{KL}[q_\phi(z | x) || p_\psi(z|y)].
\end{equation}
Different from Eq \ref{conditional_elbo}, both $q_{\phi}$ and $p_{\theta}$ in Eq \ref{relax_conditional_elbo} are not conditioned by the auxiliary covariate $y$. This allows conditional VAEs to use weights of $\phi ^ *$ and $\theta ^ *$ of a pre-trained VAE, which is trained on a diverse set of unconditional molecules, to make amortized inferences on a smaller aligned dataset of protein-ligand pairs.

%% file: results.tex
\section{Experiments}

Our implementation in PyTorch \cite{NEURIPS2019_bdbca288} and PyTorch geometric \cite{Fey:2019wv} deep learning frameworks is publicly available at \url{https://github.com/HySonLab/Ligand_Generation}.

\subsection{Binding Affinity Prediction}
\subsubsection{Experimental Setup} 
We evaluate the capability of our models on two ligand-binding datasets, DAVIS and KIBA. Our empirical results suggest that modeling long-range interactions on invariant features and leveraging sequence information provide promising performance on the task of protein-ligand affinity prediction which requires neural networks to reason on large regions of 3D structures of receptors. Both datasets contain proteins and ligands:
\begin{compactitem}
    \item \textbf{Davis}  \cite{Davis2011-fb} has 442 proteins and 68 ligands, making up 30,056 protein-ligand binding pairs, and the binding scores are measured as $K_D$ constants.
    \item \textbf{Kiba} \cite{tang2014making} has 229 proteins and 2,111 ligands, making up 118,254 protein-ligand binding pairs, and binding affinities are measured by KIBA scores.
\end{compactitem} 
Since the original datasets do not contain the three-dimensional information of the protein targets, we use data samples augmented with 3D protein structures generated by AlphaFold. These processed data are collected from the work of \citet{D3RA00281K}. For fair comparisons, we follow the same train-test split settings in \cite{ozturk2019widedta}. We use mean-squared errors (MSE), concordance index (CI), and $r_m ^ 2$ to evaluate the performance. Baseline methods include KronRLS, SimBoost, SimCNN-DTA, DeepDTA, WideDTA, AttentionDTA, MATT-DTI, GraphDTA, FusionDTA, BiCompDTA, and their results are taken from \cite{10.1371/journal.pcbi.1011036}.

\subsection{Evaluation Metrics}
We measure the performance of our models in three evaluation metrics, including mean squared error (MSE), correspondence index (CI), and $r_m ^ 2$ index.
\begin{itemize}
    \item Mean squared error (MSE):
        \begin{equation}
            \text{MSE} = \frac{1}{n} \sum_{i=1} ^ N (y_i - \hat{y_i})^2,
            \end{equation}
        where $n$ is the number of samples, $y_i$ is the observed value, and $\hat{y_j}$ is the predicted value.
    \item Concordance Index:
        \begin{equation}
    \text{CI} = \frac{1}{Z} \sum_{\delta_i > \delta_j} h(\hat{y}_i - \hat{y}_j),
        \end{equation}
    where $\hat{y}_i$ denotes the prediction for the larger affinity $\delta_i$, $\hat{y}_j$ is the predicted value for the smaller affinity $\delta_j$. Z is the normalization constant, and $h(x)$ is defined as:
    \begin{equation}
    h(x) = 
    \begin{cases}
        1 & \quad x > 0 \\
        0.5 & \quad x = 0 \\
        0 & \quad  x< 0
     \end{cases}
    \end{equation}
    
    \item $r_m ^ 2$ Index:    
    \begin{equation}
        r_m ^ 2 = r^2 \times \bigg(1 - \sqrt{r^2 - r_0^2} \bigg),
    \end{equation}
    where $r^2$ and $r_0 ^ 2$ are the squared correlation coefficients with and without intercepts respectively.
\end{itemize}

\subsubsection{Experimental Results}
We conduct a five-fold validation (given in the dataset) to select the optimal weights for PMN. According to Table \ref{tab:results_davis_kiba}, our method outperforms the baselines on the DAVIS dataset by a large margin and achieves comparable performance to the best competitor on the KIBA dataset. Similar to PMN, FusionDTA also augments the representations of proteins by adopting ESM-1b \cite{rives2021biological} Transformer encoder for producing representation vectors of protein sequences. However, instead of leveraging efficient Transformers like ours, the approach utilizes full-rank Transformers for learning on long protein sequences, which requires excessive computational resources for training and fine-tuning. This can explain the trade-off between performance and training efficiency between our method and FusionDTA. On the other hand, while BiCompDTA demands carefully-processed features to encode protein sequences, our approach can learn this information directly from raw structures and sequences in a data-driven manner. 

\begin{table}
\small
    \caption{Experimental Results on DAVIS and KIBA dataset. Results are averaged over five runs.}
    \label{tab:results_davis_kiba}
    \vspace{5pt}
    \centering
    \begin{tabular}{lcccccc} 
         \toprule  
         \multirow{2}{*}{Approach} & \multicolumn{3}{c}{DAVIS} & \multicolumn{3}{c}{KIBA} \\
         \cmidrule{2-7} 
         & MSE $\downarrow$ & CI $\uparrow$ & $r_m ^ 2 \uparrow$ & MSE $\downarrow$ & CI $\uparrow$ & $r_m ^ 2 \uparrow$ \\
         \midrule
         KronRLS \cite{nascimento2016multiple} & 0.379 & 0.871 & 0.407 & 0.411 & 0.782 & 0.342 \\
         SimBoost \cite{He2017} & 0.282 & 0.872 & 0.644 & 0.222 & 0.836 & 0.629 \\
         SimCNN-DTA \cite{shim2021prediction} & 0.319 & 0.852 & 0.595 & 0.274 & 0.821 & 0.573 \\
         DeepDTA \cite{10.1093/bioinformatics/bty593} & 0.261 & 0.878 & 0.63 & 0.194 & 0.863 & 0.673 \\
         WideDTA \cite{ozturk2019widedta} & 0.886 & 0.262 & --- & 0.875 & 0.179 & --- \\
         AttentionDTA \cite{zhao2019attentiondta} & 0.216 & 0.893 & 0.677 & 0.155 & \underline{0.882} & 0.755 \\
         MATT-DTI \cite{10.1093/bib/bbab117} & 0.227 & 0.891 & 0.683 & \textbf{0.150} & \underline{0.882}  & 0.756 \\
         GraphDTA \cite{D3RA00281K} & 0.258 & 0.884 & 0.656 & 0.162 & 0.879 & 0.736 \\
         FusionDTA \cite{10.1093/bib/bbab506} & 0.220 & 0.903 & 0.666 & 0.167 & \textbf{0.891} & 0.699 \\
         BiCompDTA \cite{10.1371/journal.pcbi.1011036} & 0.237 & 0.904 & 0.696 & 0.167 & \textbf{0.891} & \underline{0.757} \\
         \midrule
         PMN (ours) & \textbf{0.202} & \textbf{0.906} & \textbf{0.739} & \underline{0.153} &
0.874  & \textbf{0.767} \\
std & $\pm$ 0.007 & $\pm$ 0.003 & $\pm$ 0.011 & $\pm$ 0.002 & $\pm$ 0.003 & $\pm$ 0.003
 \\
         \bottomrule
    \end{tabular}
\end{table}

\begin{figure}
    \captionsetup[subfigure]{labelformat=empty}
    \centering
    \subfloat[Fold 1]{
    \includegraphics[width=7.5cm]{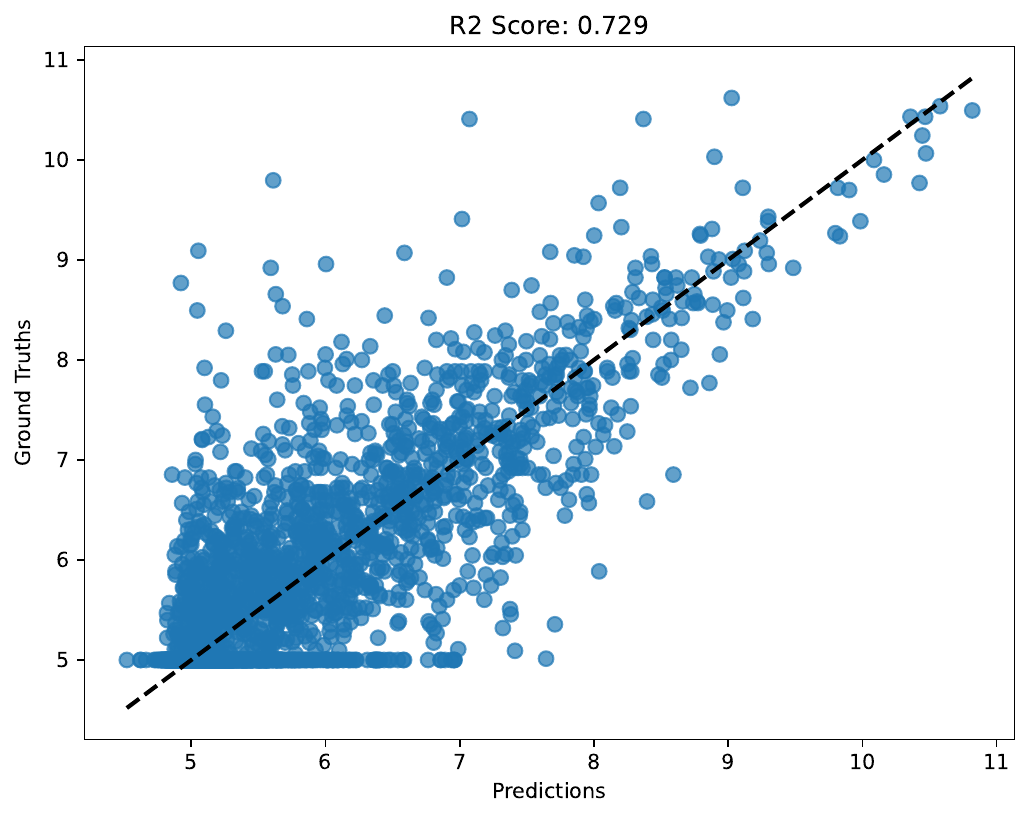}
    }
    \subfloat[Fold 2]{
    \includegraphics[width=7.5cm]{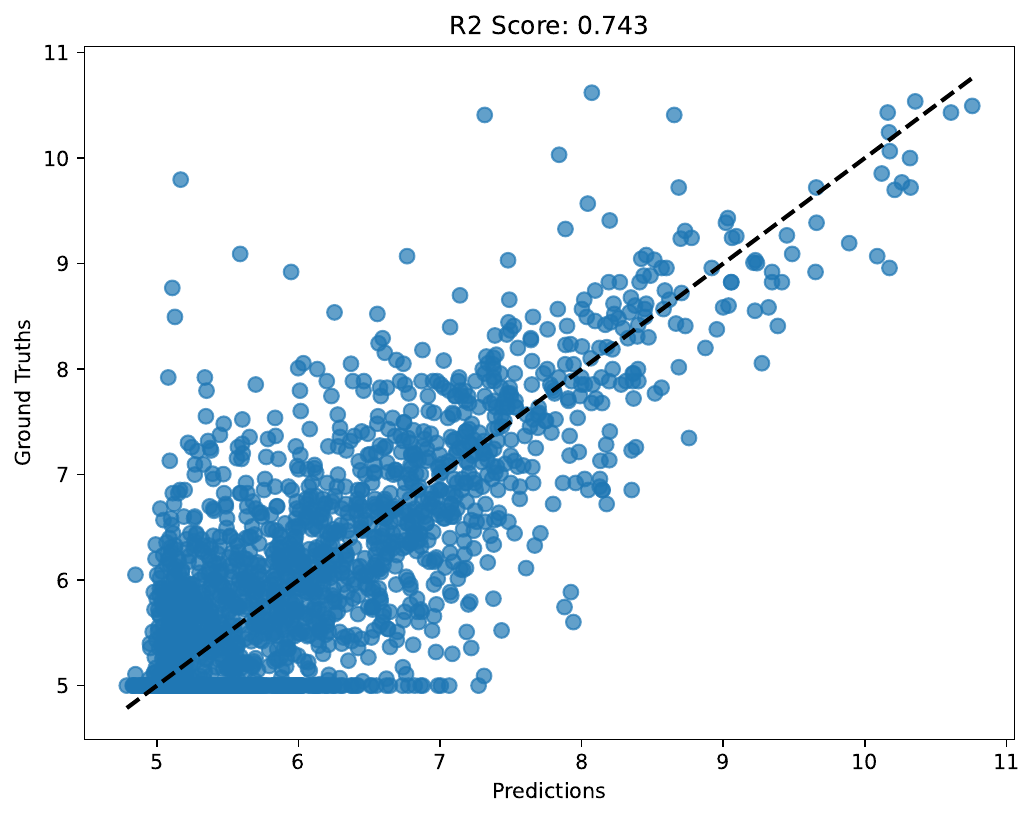}
    }
    \quad
    \subfloat[Fold 3]{
    \includegraphics[width=7.5cm]{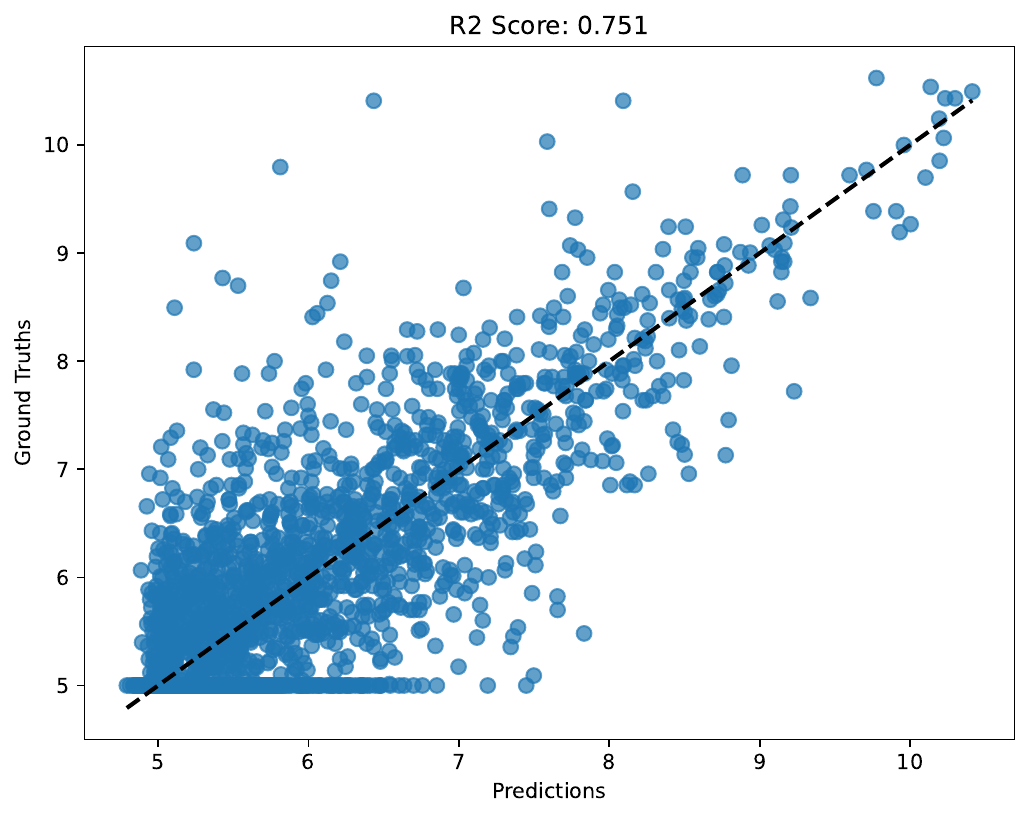}
    }
    \subfloat[Fold 4]{
    \includegraphics[width=8.5cm]{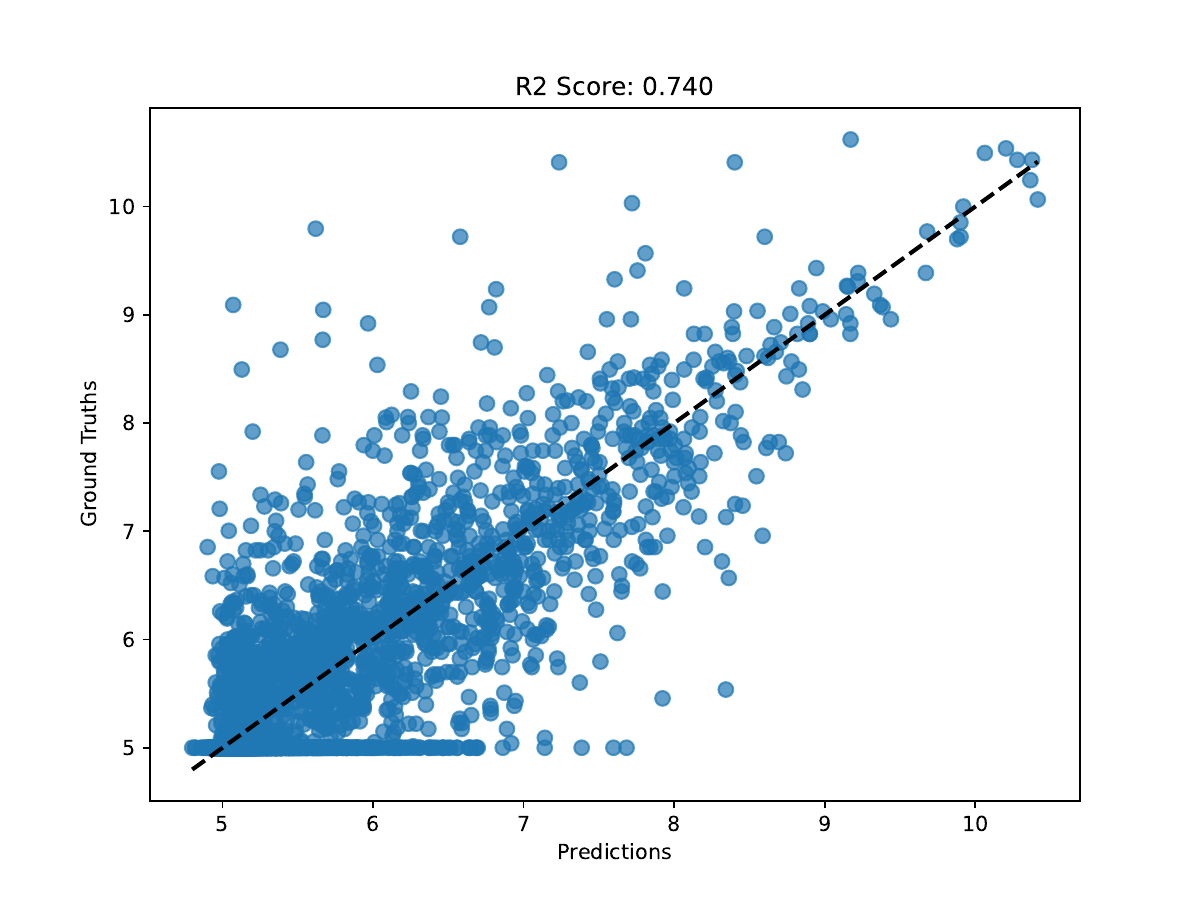}
    }
    \quad
    \subfloat[Fold 5]{
    \includegraphics[width=7.5cm]{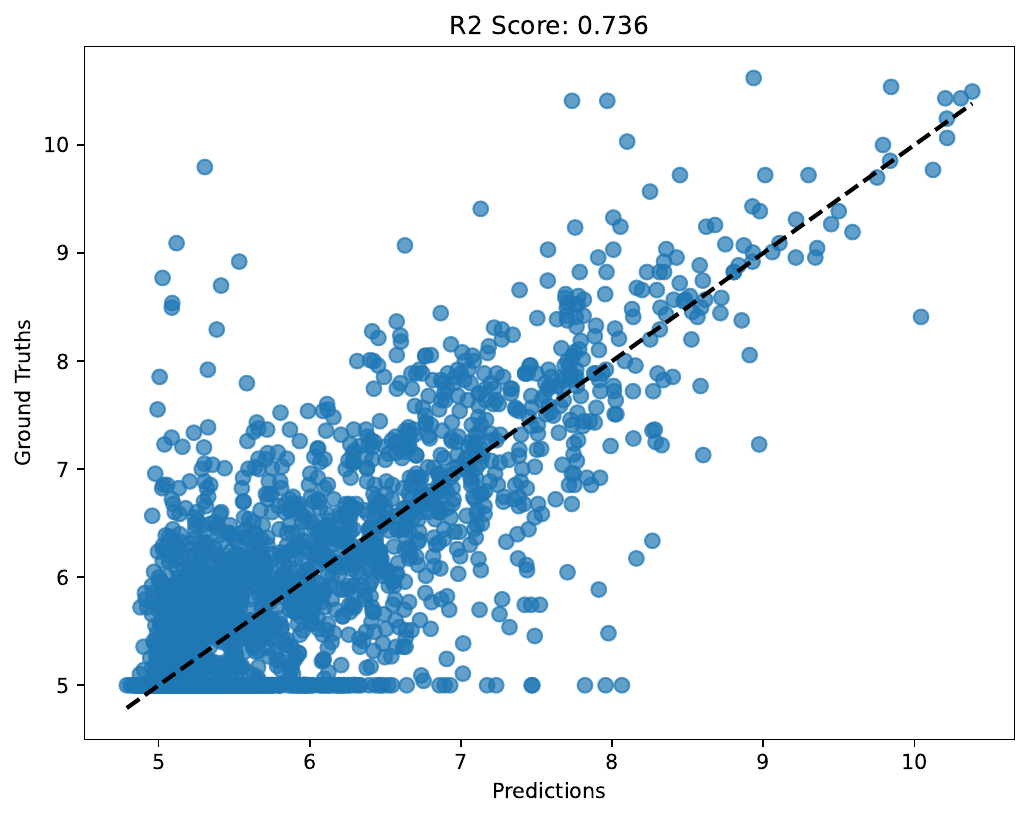}
    }
    \caption{$R^2$-scores of our models trained in five different folds of the DAVIS dataset, indicating the goodness of fit of our proposed method to unseen protein-ligand complexes in the testing sets.}
    \label{fig:r2_score_davis}
\end{figure}

\begin{figure}
    \captionsetup[subfigure]{labelformat=empty}
    \centering
    \subfloat[Fold 1]{
    \includegraphics[width=8cm]{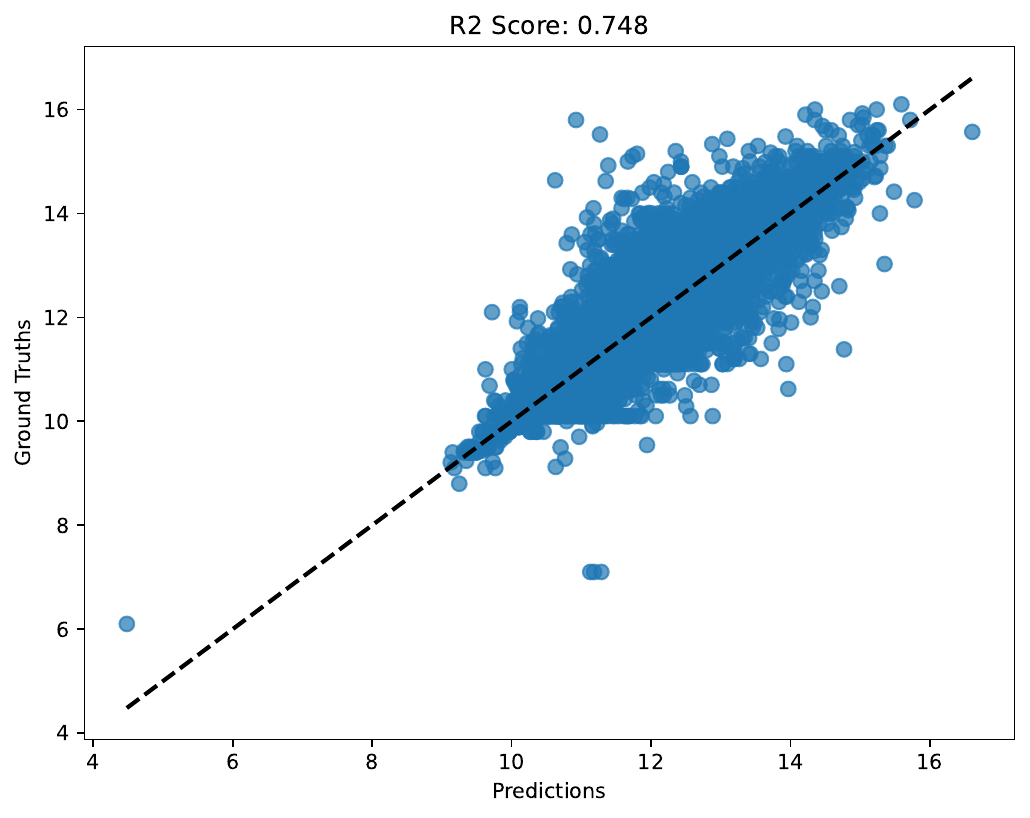}
    }
    \subfloat[Fold 2]{
    \includegraphics[width=8cm]{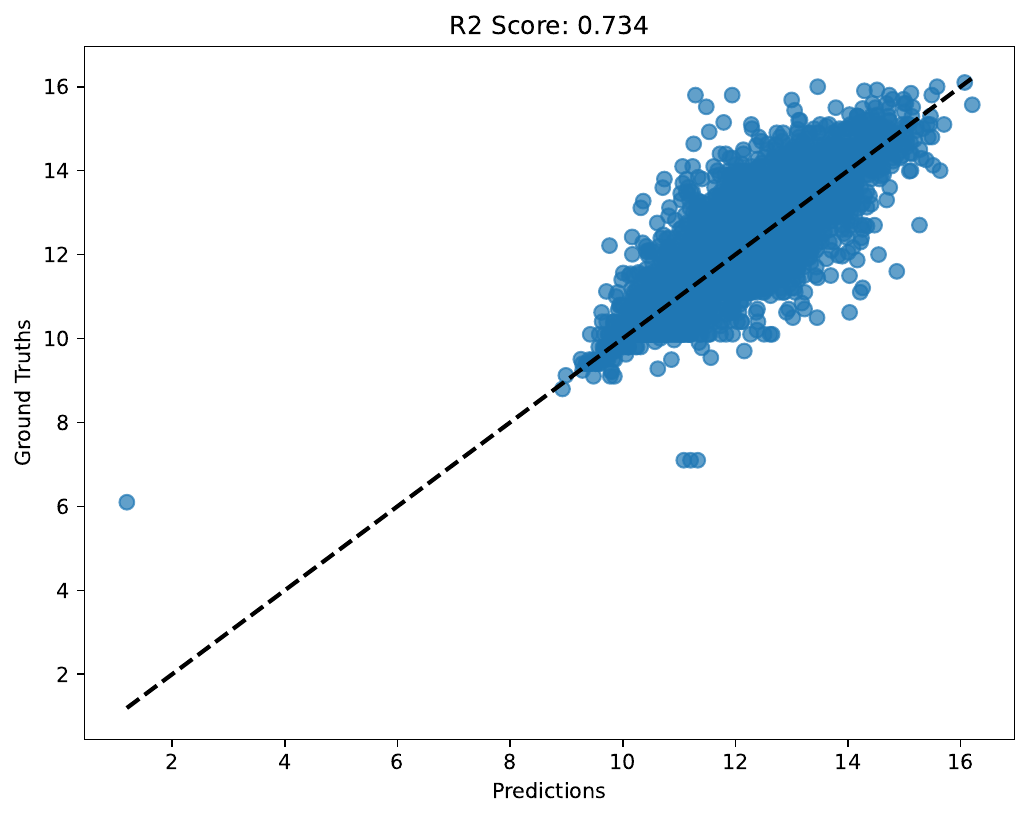}
    }
    \quad
    \subfloat[Fold 3]{
    \includegraphics[width=8cm]{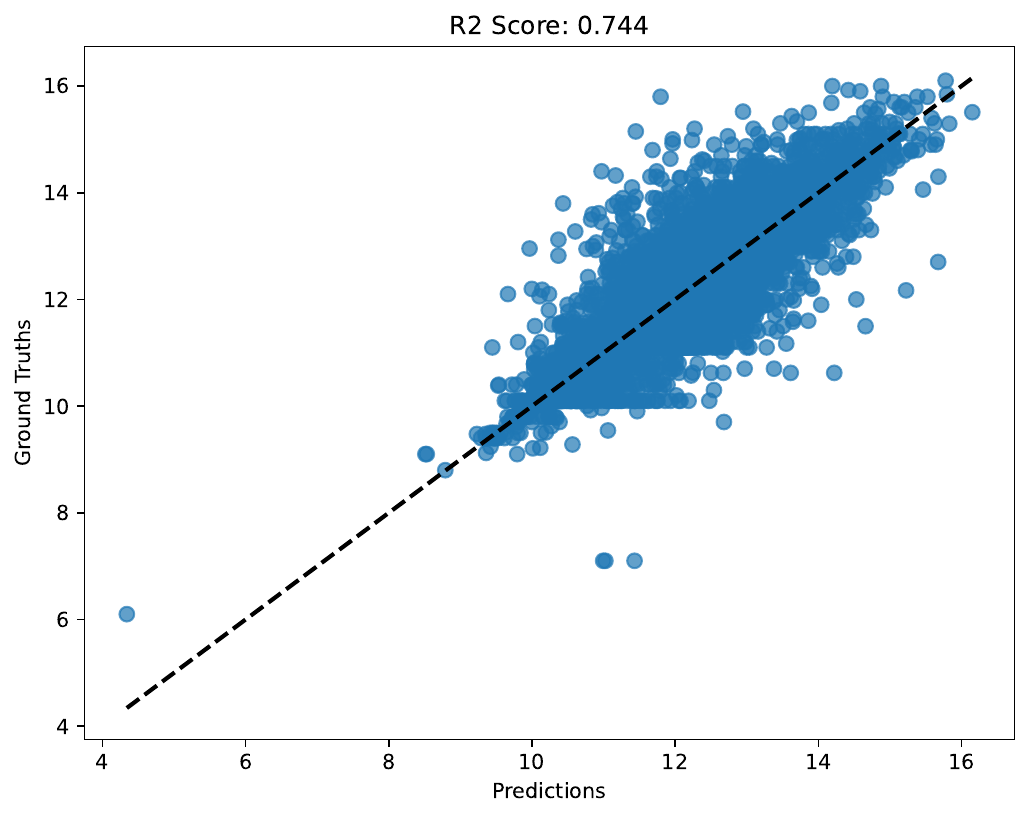}
    }
    \subfloat[Fold 4]{
    \includegraphics[width=8cm]{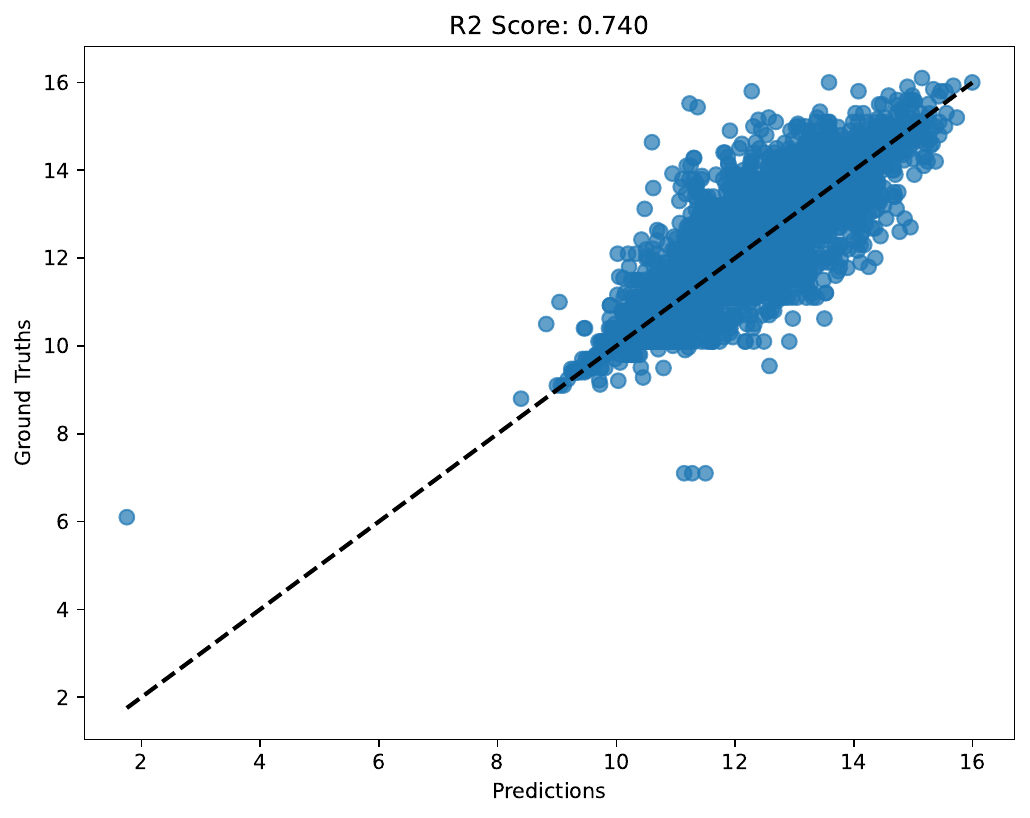}
    }
    \quad
    \subfloat[Fold 5]{
    \includegraphics[width=8cm]{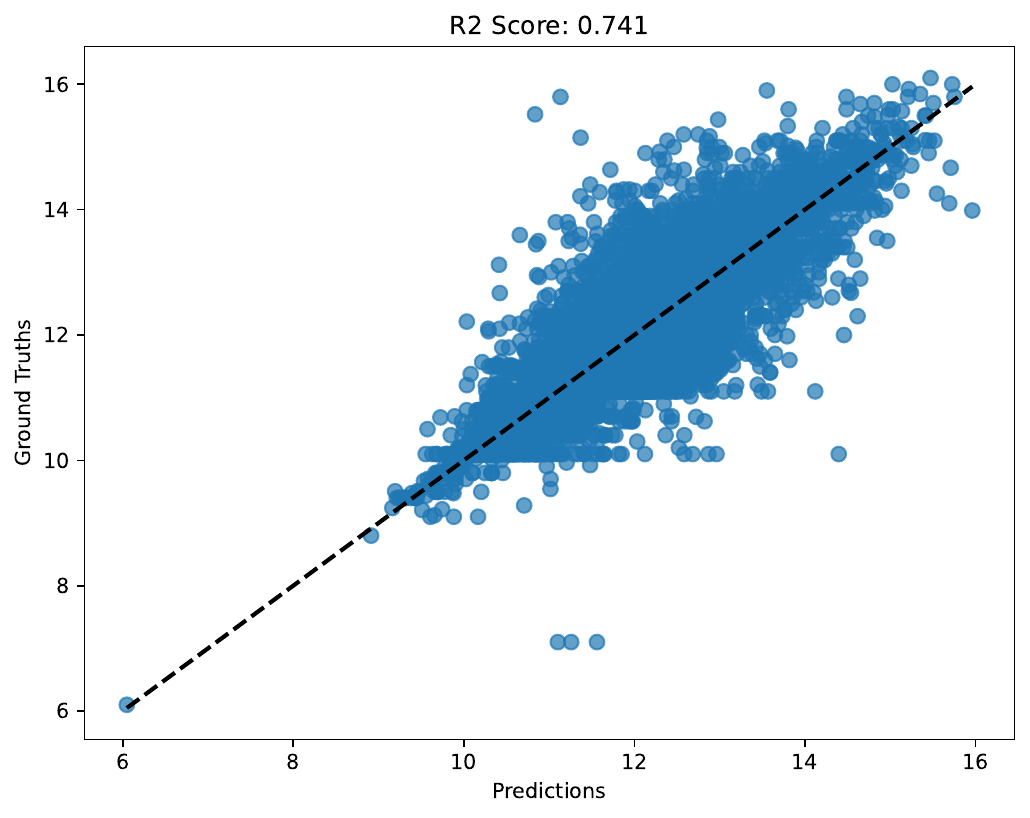}
    }
    \caption{$R^2$-scores of our model trained in five different folds of the KIBA dataset, indicating the goodness of fit of our proposed method to unseen protein-ligand complexes in the testing sets.}
    \label{fig:r2_score_kiba}
\end{figure}



\subsection{Target-aware Drug Design}

\subsubsection{Experimental Setup}
\paragraph{Dataset} We utilize the dataset KIBA \cite{tang2014making} for conditional molecule generation. KIBA contains 229 proteins and 2,111 ligands, and there are 118,254 protein-ligand pairs in total. For unconditional pre-training, we train a VAE on the ZINC250K dataset, which contains about 250,000 drug-like molecules. As both datasets provide SMILES as representations for the molecules, we convert them to SELFIES representations. We filter out SMILES that can not be converted to SELFIES and build a vocabulary of SELFIES blocks, which consists of 108 tokens. We split the dataset into 90 $\%$ of proteins for training and $10\%$ of targets for testing.

\paragraph{Quality of generative models} We use the Fréchet ChemNet Distance (FCD) \cite{Preuer2018} score to identify the quality of generative models. FCD measures the distance between the distribution of real-world molecules and the distribution of the generated molecules. A well-performing generative model should have a low FCD score, indicating that it can approximate the distribution of real-world molecules. In addition to FCD, we measure three biological and chemical properties of the generated molecules, including drug-likeness (QED), penalized lipophilicity (pLogP), and synthetic accessibility (SA). We generate 1,000 ligands for each target in the testing dataset, resulting in 23,000 generated molecules in total. Then, considering the specified protein target as a condition, we calculate the distances between the generated molecules and real-world compounds (i.e. ligands that pair with the target in the dataset).

\paragraph{Docking Measurements} We use the AutoDock-GPU software to compute the binding affinity scores of the generated ligands to protein targets. We test the performance of our conditional model on two popular binding sites, ESR1 (UniProt: P03372) and ACAA1 (UniProt: P09110). Notably, both targets are unseen in the training phase and not included in KIBA. The docking scores are measured by  dissociation constants, $K_D$, in nanomoles/liter (lower means higher binding affinities). To the best of our knowledge, our work is the first to introduce generative models conditioned on arbitrary protein targets. As a result, comparing it with existing methods optimized for specific targets becomes challenging. For comparisons, we select methods that are optimized for ESR1 and ACAA1 and take their results from \cite{eckmann2022limo}. In particular, we generate 10,000 molecules for each target and use AutoDock-GPU software for docking computation.

\subsubsection{Experimental Results}

\begin{figure}
    \centering
    \subfloat[Target]{
    \includegraphics[scale = 0.15]{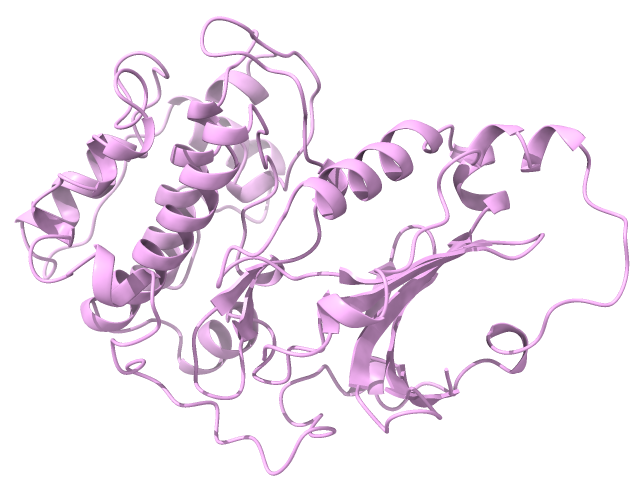}
    }
    \subfloat[$O_\text{for}$]{
    \includegraphics[scale = 0.3]{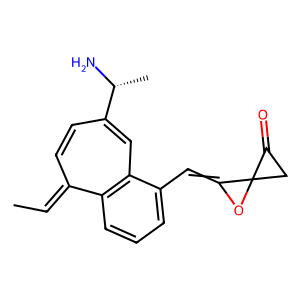}
    }
    \subfloat[$O_\text{for}$]{
    \includegraphics[scale = 0.3]{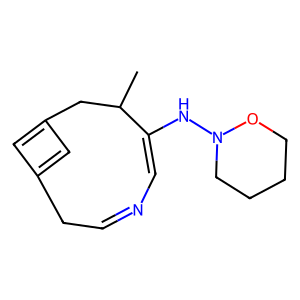}
    }
    \subfloat[$O_\text{cond}$]{
    \includegraphics[scale = 0.3]{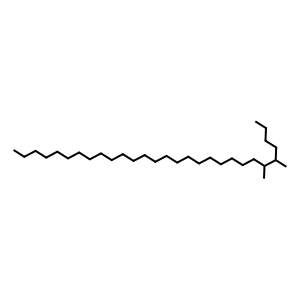}
    }
    \caption{2D illustration of molecules generated by TargetVAE trained with two objectives $O_\text{for}$ and $O_\text{cond}$}
    \label{fig:2D_visualization}
\end{figure}

\begin{table}
\caption{Comparison between $O_\text{cond}$ and $O_\text{for}$}\label{tab:qed} 
\centering
\begin{tabular}{cccc} \\
\toprule  
Objective & QED $\uparrow$ & SA $\downarrow$ & pLogP $\uparrow$ \\\midrule
$O_\text{cond}$ & 0.118 & 2.48 & 9.57 \\  \midrule
$O_\text{for}$ & 0.913 & 1.29 & 3.81\\ 
\bottomrule
\end{tabular}
\end{table} 

\begin{figure}[h]
    \centering
    \includegraphics[scale = 1.0]{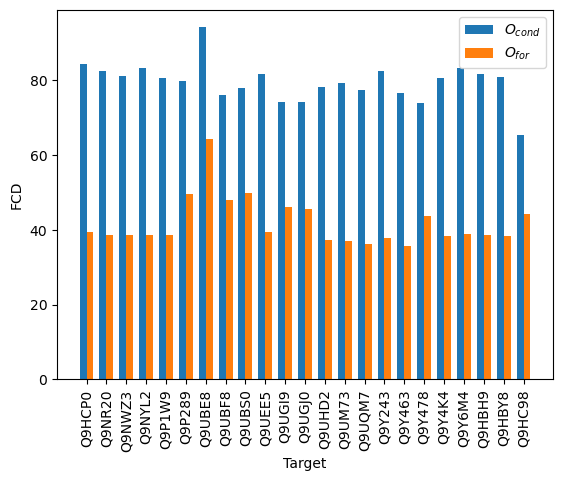}
    \caption{Average FCD scores corresponding to protein targets in the test dataset}
    \label{fig:fcd}
\end{figure}

\paragraph{Approximation of real distributions} This experiment aims to explore the capabilities of conditional VAE in generating real-world molecules given their corresponding targets. Figure \ref{fig:fcd} illustrates the FCD scores of TargetVAE trained with objectives $O_\text{cond}$ in Eq. \ref{conditional_elbo} and $O_\text{for}$ in Eq. \ref{relax_conditional_elbo}. For each target, lower FCD scores show that TargetVAE trained by $O_\text{for}$ approximate the distributions of real-world molecules better than that trained by $O_\text{cond}$. We further explore this phenomenon by visualizing the generated molecules and recognize that posterior collapse happens when TargetVAE is trained by $O_\text{cond}$. According to Figure \ref{fig:2D_visualization}, samples generated by TargetVAE trained by $O_\text{cond}$ collapse to simple molecules, while the model trained by $O_{for}$ can generate diverse samples for each protein target. Moreover, Table \ref{tab:qed} shows the average scores of the top ten molecules in three properties (i.e. QED, SA, and pLogP). In particular, TargetVAE trained by $O_\text{for}$ can generate molecules with high QED (>90) and low SA (<2), which are uncommon in drug discovery \cite{bickerton2012quantifying}. In contrast, the model trained by $O_\text{cond}$ converges to molecules with high pLogP values, while possesing very low QED, indicating that they are not drug-like molecules.

\begin{figure}
\centering
\includegraphics[width=0.45\linewidth]{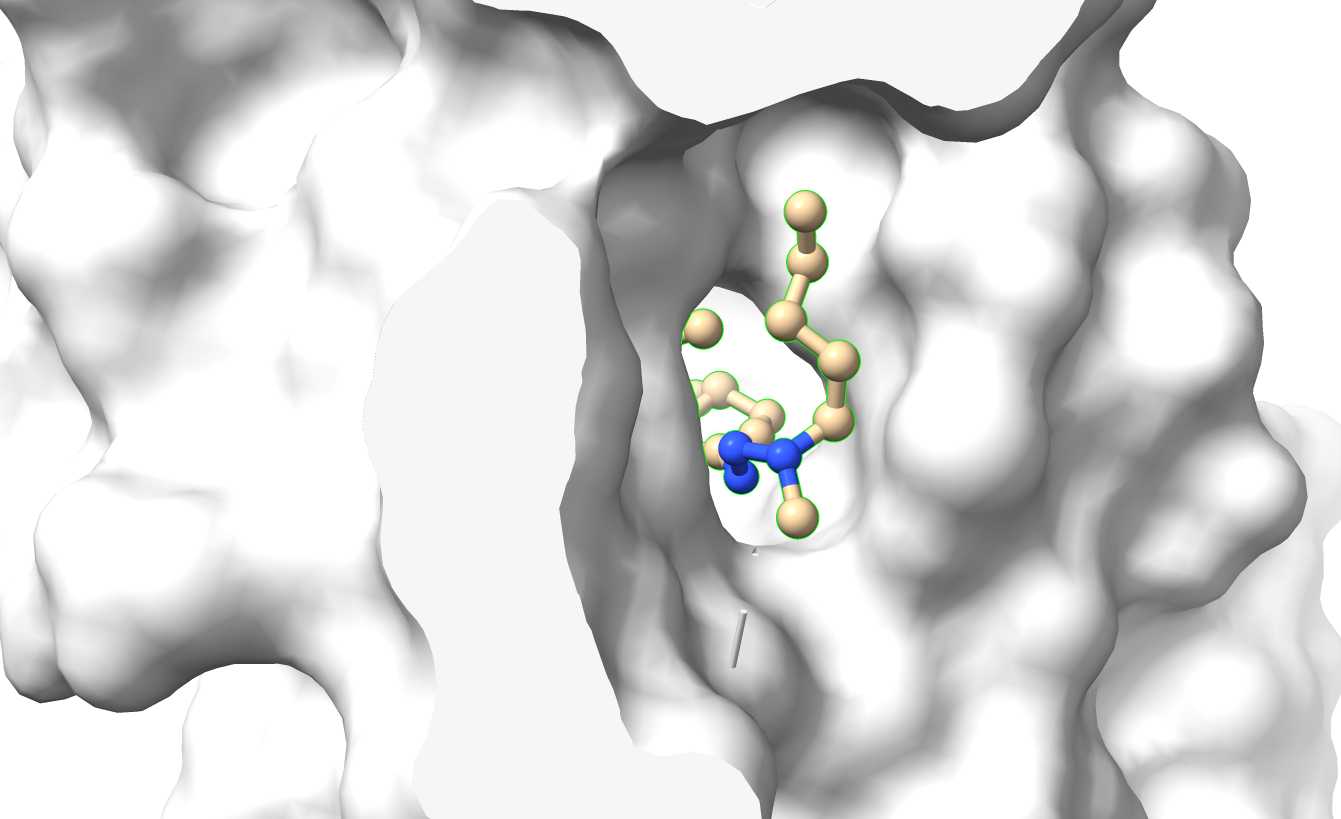}
\includegraphics[width=0.45\linewidth]{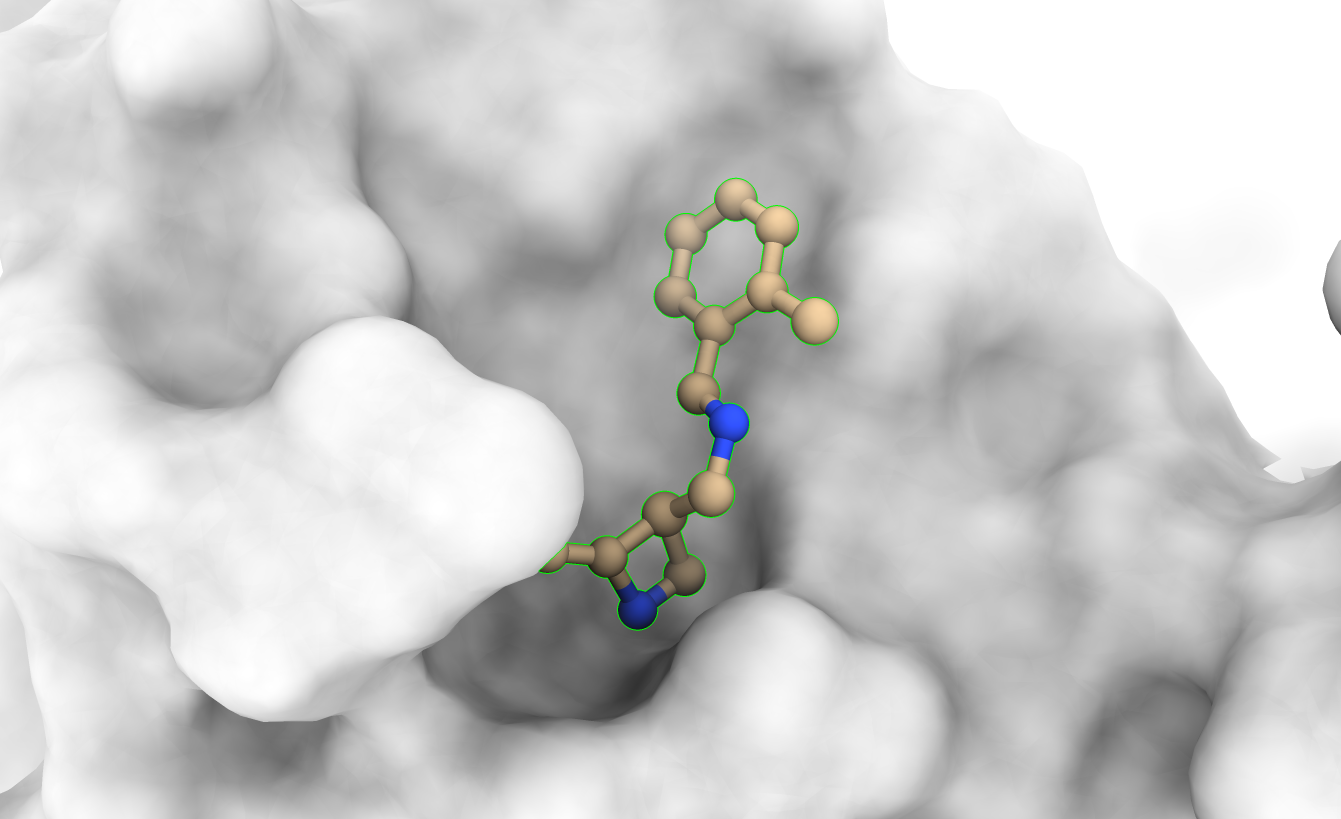}
\\
\includegraphics[width=0.45\linewidth]{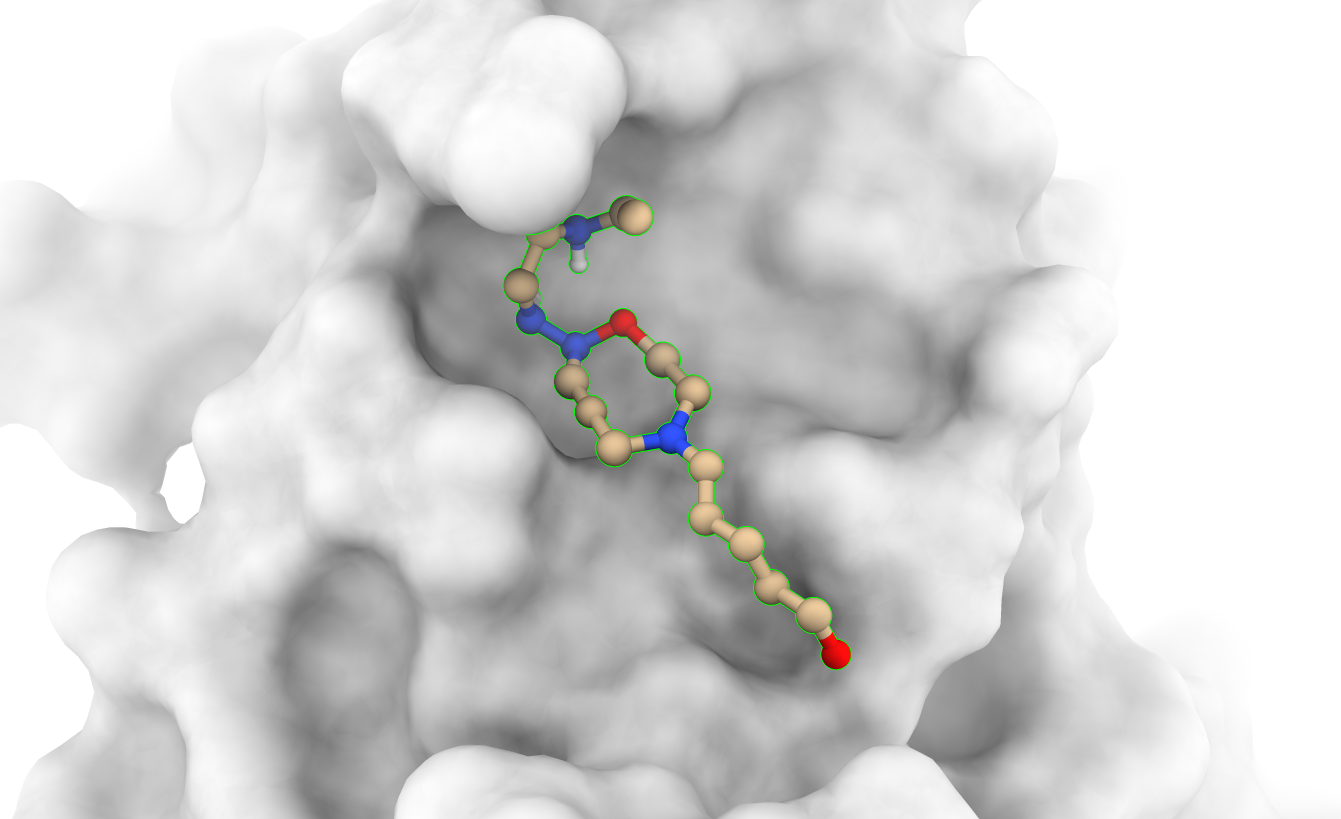}
\includegraphics[width=0.45\linewidth]{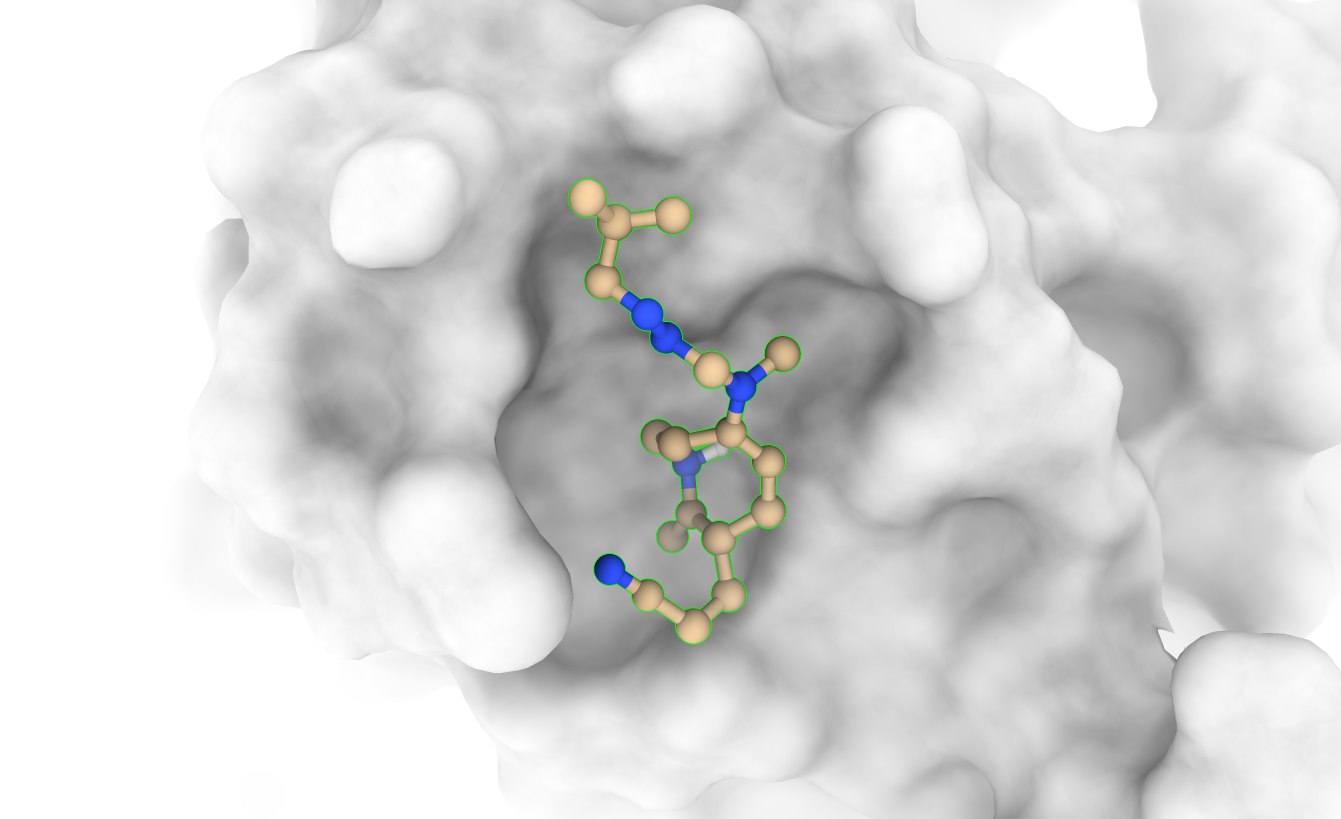}
\caption{3D visualizations of how ligands bind to ERR1. 3D conformations of generated molecules are calculated by the Obabel software \cite{o2011open}.}
\label{fig:3D_binding_err1}
\end{figure}

\begin{figure}
\centering
\includegraphics[width=0.45\linewidth]{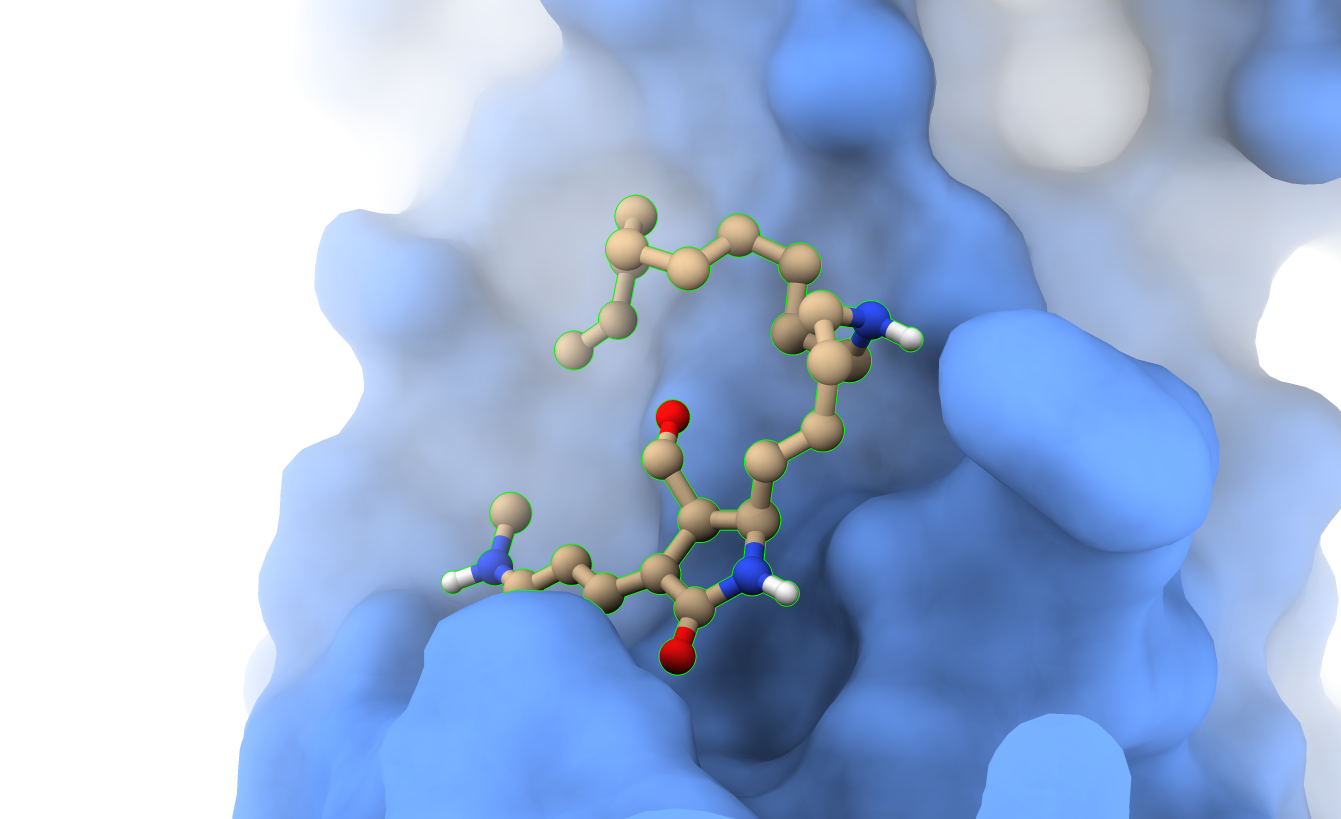}
\includegraphics[width=0.45\linewidth]{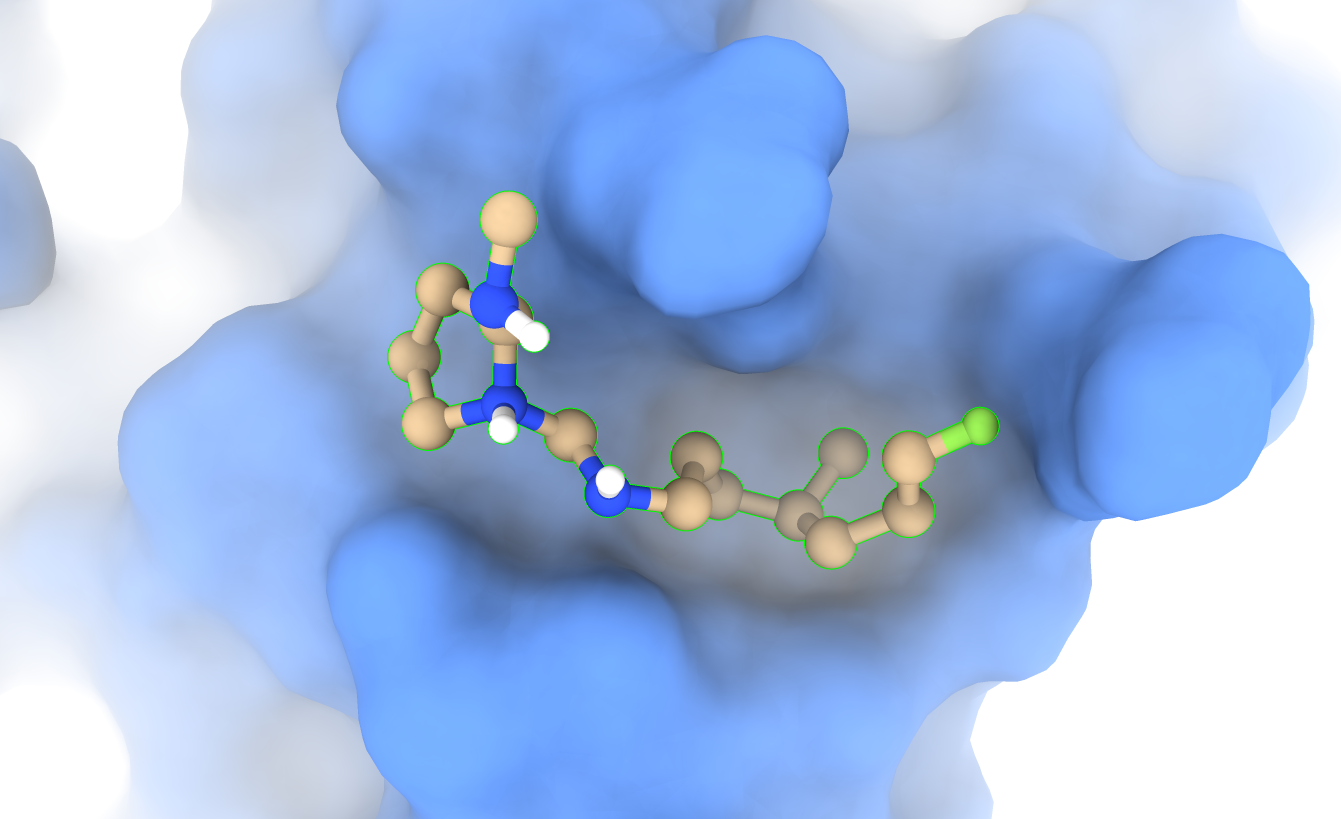}
\\
\includegraphics[width=0.45\linewidth]{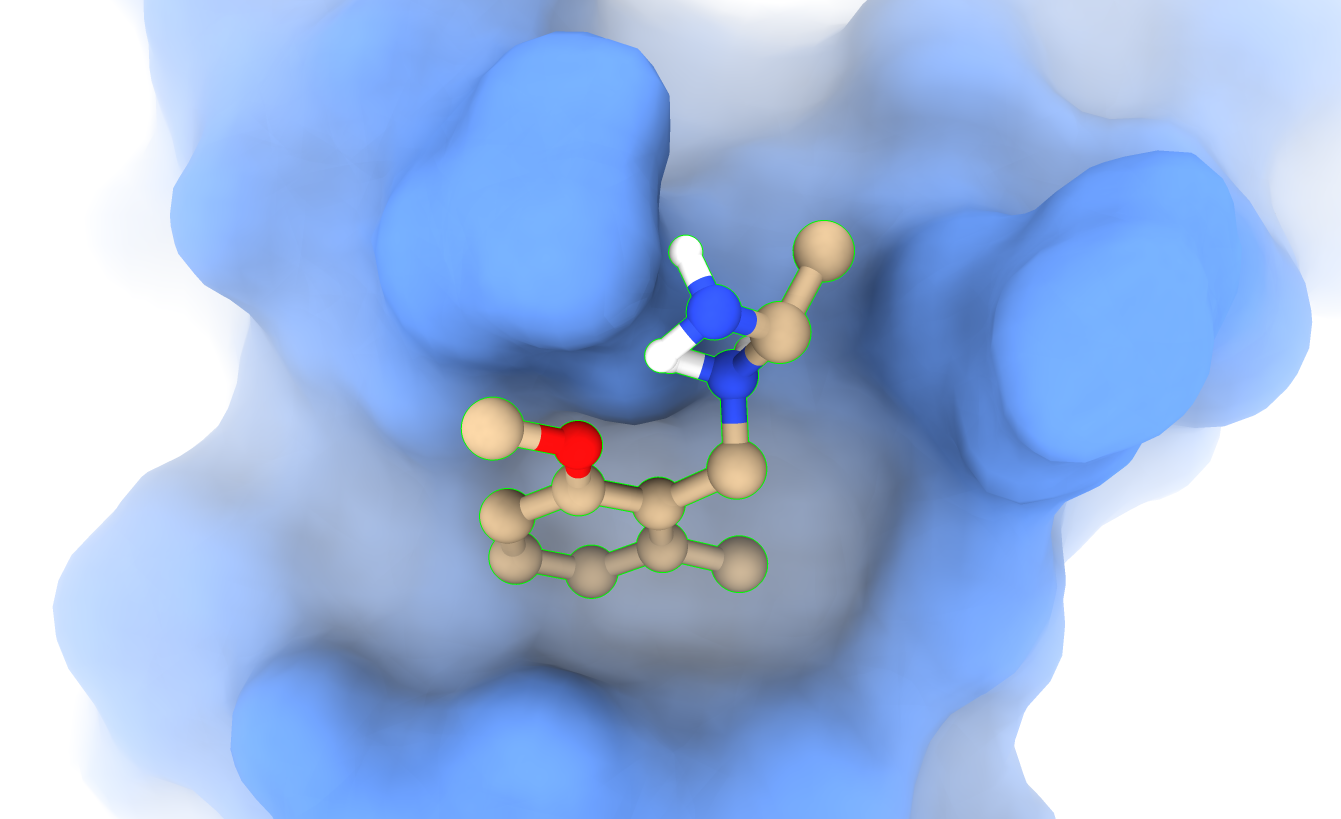}
\includegraphics[width=0.45\linewidth]{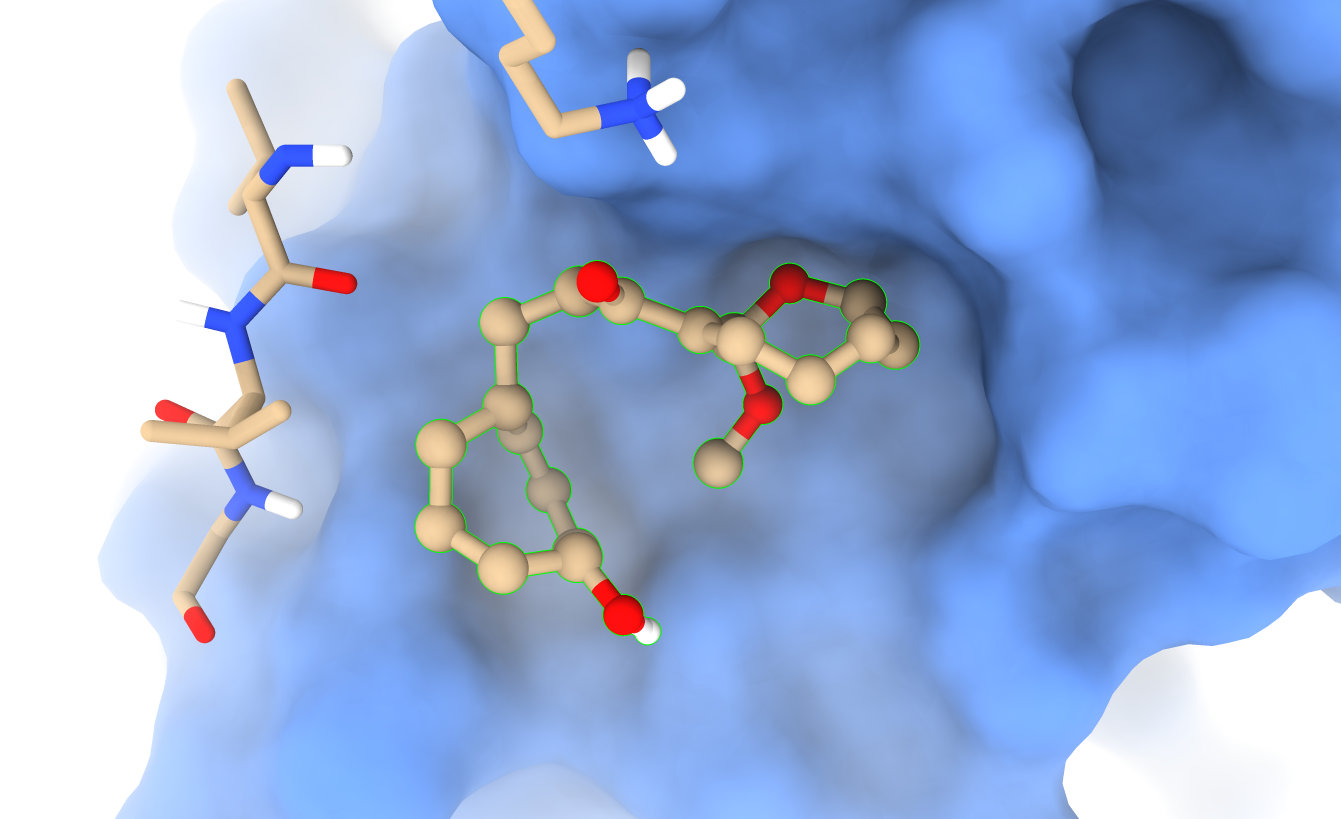}
\caption{3D visualizations of how ligands bind to ACC1. 3D conformations of generated molecules are calculated by the Obabel software \cite{o2011open}.}
\label{fig:3D_binding_acc1}
\end{figure}

\begin{figure}
    \captionsetup[subfigure]{labelformat=empty}
     \centering 
     \subfloat[][Q9P289]{
     \includegraphics[scale = 0.12]{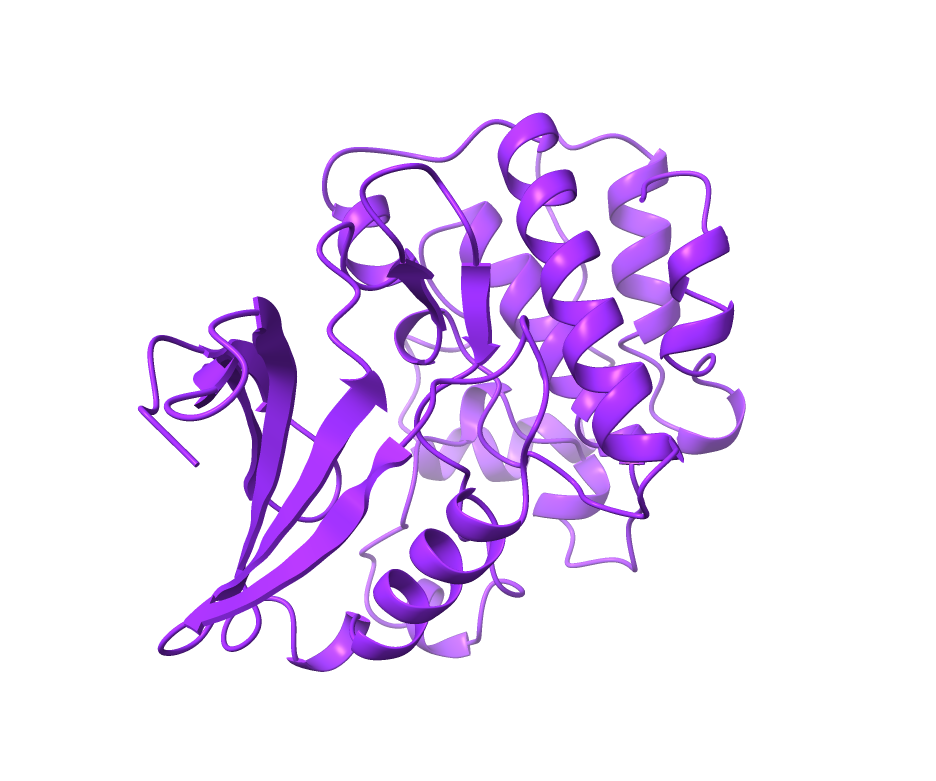}
     }
     \subfloat[QED: 0.9314]{
     \includegraphics[scale = 0.36]{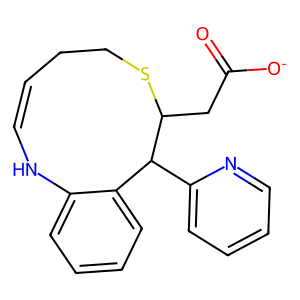}
     }
     \subfloat[QED: 0.922]{
     \includegraphics[scale = 0.36]{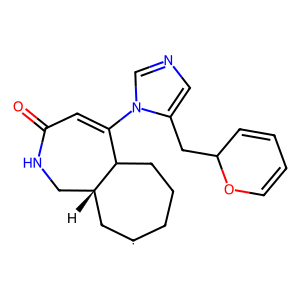}
     }
     \subfloat[QED: 0.887]{
     \includegraphics[scale = 0.36]{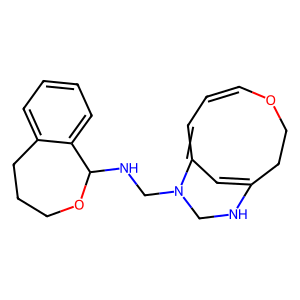}
     }
     \quad
     \subfloat[][Q9HBH9]{
     \includegraphics[scale = 0.15]{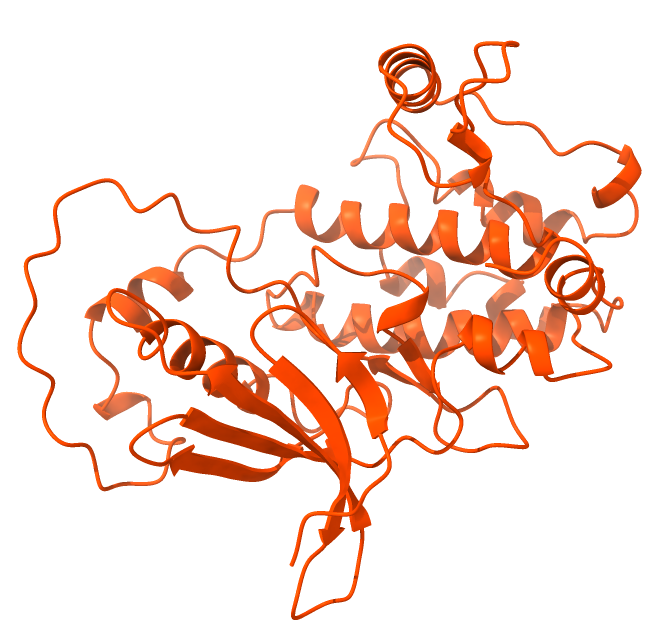}
     }
     \subfloat[][QED: 0.867]{
     \includegraphics[scale = 0.36]{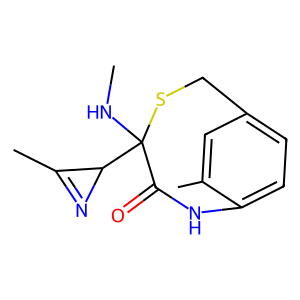}
     }
    \subfloat[][QED: 0.843]{
    \includegraphics[scale = 0.36]{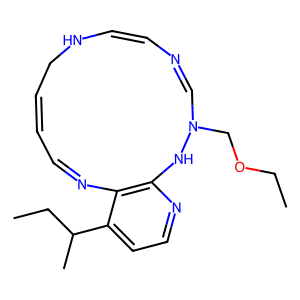}
    }
    \subfloat[][QED: 0.907]{
    \includegraphics[scale = 0.36]{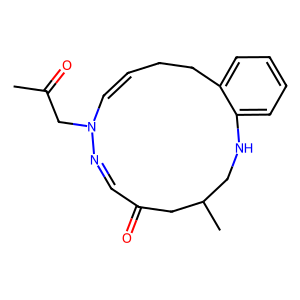}
    }
    \quad 
    \subfloat[][Q9Y6M4]{
    \includegraphics[scale = 0.09]{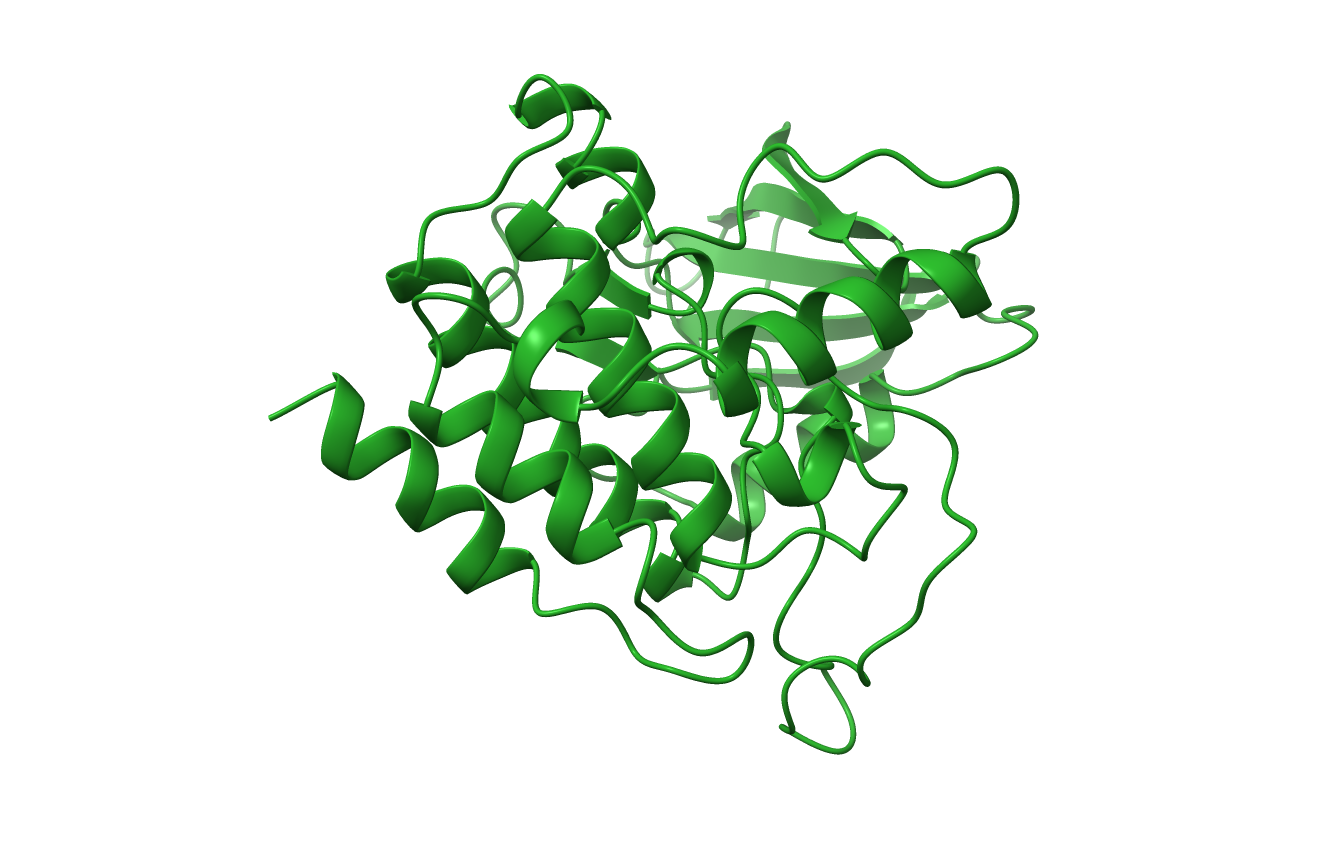}
    }
    \subfloat[][QED: 0.905]{
    \includegraphics[scale = 0.36]{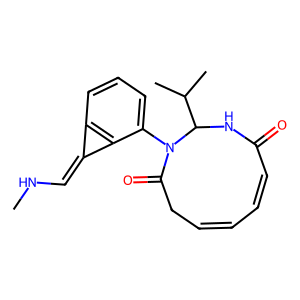}
    }
    \subfloat[][QED: 0.870]{
    \includegraphics[scale = 0.36]{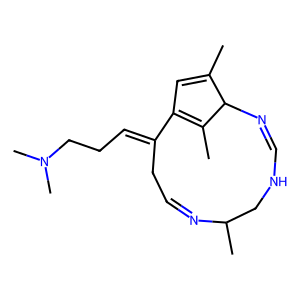}
    }
    \subfloat[][QED: 0.858]{
    \includegraphics[scale = 0.36]{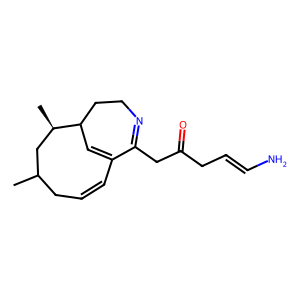}
    }
    \quad 
    \subfloat[][Q9UEE5]{
    \includegraphics[scale = 0.09]{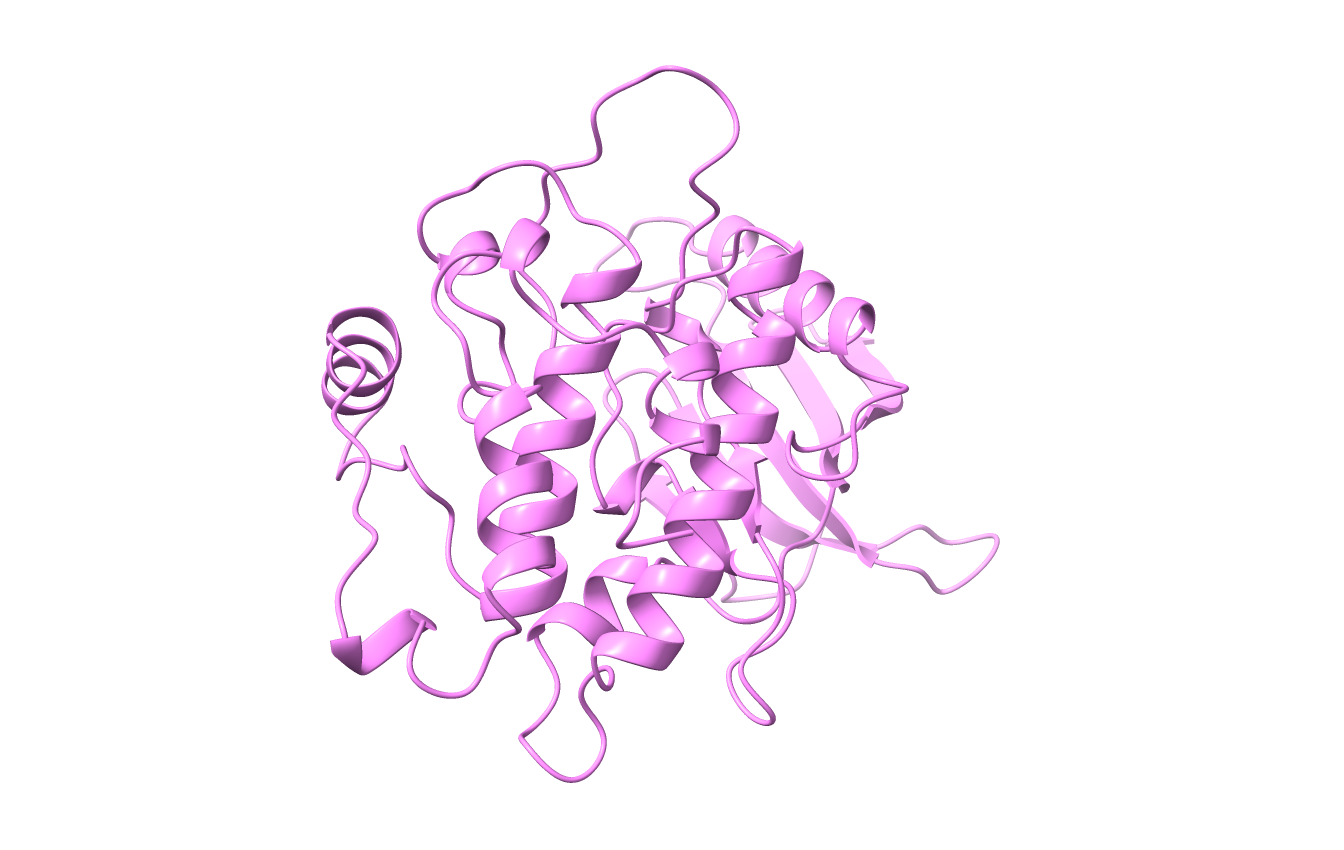}
    }
    \subfloat[][QED: 0.904]{
    \includegraphics[scale = 0.36]{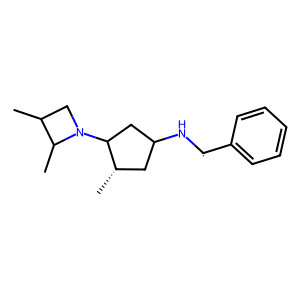}
    }
    \subfloat[][QED: 0.856]{
    \includegraphics[scale = 0.36]{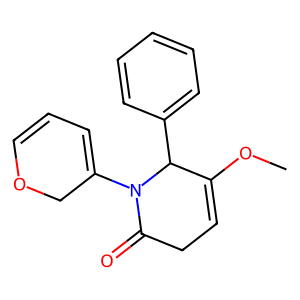}
    }
    \subfloat[][QED: 0.847]{
    \includegraphics[scale = 0.36]{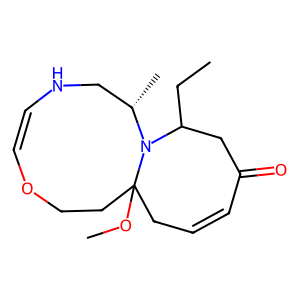}
    }
    \caption{\label{fig:high-qed} Visualizations of ligands generated by TargetVAE, given their protein targets with their UniProt IDs. Our generated ligands have high drug-likeness scores (i.e. QED).}
\end{figure}

\paragraph{Zero-shot generation to arbitrary targets}
As shown in Table \ref{results:top_3}, TargetVAE can generate molecules with higher binding affinities (lower $K_D$) than prior state-of-the-art RL-based or iterative methods like GCPN, MOLDQN, GraphDF, and MARS. In this setup, TargetVAE is trained with the $O_\text{for}$ objective in Eq. \ref{relax_conditional_elbo}. We achieve comparable performance with LIMO, another VAE-based approach. While effective, LIMO shares a similarity with RL-based methods in that it requires training a specific property network for each protein target, resulting in inefficiency and limitations when dealing with a large number of targets. Moreover, optimizing molecules for high binding affinities may compromise other critical properties such as QED and SA, leading to sub-optimal overall performance. To prove this fact, we select two molecules that have the highest QED scores and make comparisons with those produced by LIMO and GCPN in \cite{eckmann2022limo}. Table \ref{tab:experimental_binding_2} demonstrates that while having the lowest $K_D$, ligands generated by LIMO are not likely drug-like molecules. In contrast, GCPN can generate drug-like molecules with QED up to 0.80, yet the method fails at producing ligands with high binding affinities. TargetVAE, on the other hand, offers the advantage of maintaining a balance among properties. Our method excels at generating ligands that possess desirable drug-like qualities, and synthetic accessibility, while still exhibiting reasonably favorable binding affinities. Finally, Figures \ref{fig:3D_binding_err1} and \ref{fig:3D_binding_acc1} show how the generated ligands bind to their targets. Figure \ref{fig:high-qed} visualizes some of our generated ligands with high drug-likeness scores (i.e. QED) along with their target proteins.


\begin{table}[h]
\small
  \caption{Top-three generated molecules with high binding affinities (shown as $K_D \downarrow$) for ESR1 and ACAA1.}
  \label{results:top_3}
  \centering
  \begin{tabular}{lcccccc} 
    \toprule 
    \multirow{2}{*}{Method} & \multicolumn{3}{c}{ESR1} & \multicolumn{3}{c}{ACAA1} \\
    \cmidrule(lr){2-7} 
           &  1ST &  2ND & 3RD & 1ST & 2ND & 3RD \\
    \midrule 
    GCPN \cite{NEURIPS2018_d60678e8}  &  6.4 &  6.6 & 8.5 & 75  & 83  & 84 \\ 
    MOLDQN \cite{zhou2019optimization} &  373 &  588 & 1062 & 240 & 337 & 608 \\
    GraphDF \cite{pmlr-v139-luo21a} & 25 & 47 & 51 & 370 & 520 & 590 \\
    MARS \cite{xie2021mars} & 17 & 64 & 69 & 163 & 203 & 236  \\
    LIMO \cite{eckmann2022limo}& 0.72 & 0.89 & 1.4 & 37 & 37 & 41 \\ 
    \midrule
     TargetVAE (ours) & 0.55 & 2.7 & 5.1 & 87.3 & 165 & 177 \\
    \bottomrule
  \end{tabular}
\end{table}

\begin{table}[h]
    \centering
    \begin{tabular}{lcccccc} \\
         \toprule
         \multirow{2}{*}{Ligand} & \multicolumn{3}{c}{ESR1} & \multicolumn{3}{c}{ACAA1} \\
        \cmidrule(lr){2-7} 
          & $K_D \downarrow$ & QED $\uparrow$ & SA $\downarrow$ & $K_D (\downarrow)$ & QED $\uparrow$ & SA $\downarrow$ \\
         \midrule
         LIMO $\#$1 & 4.6 & 0.43 & 4.8 & 28 & 0.57 & 5.5 \\
         LIMO $\#$2 & 2.8 & 0.64 & 4.9 & 31 & 0.44 & 4.9 \\
         \midrule 
         GCPN $\#$1 & 810 & 0.43 & 4.2 & 8500 & 0.69 & 4.2 \\
         GCPN $\#$2 & 2.7 $\times 10^4$ & 0.80 & 3.7 & 8500 & 0.54 & 4.2 \\
         \midrule 
         TargetVAE (ours) $\#$ 1 & 100 & 0.79 & 6.0 & 420 & 0.77 & 5.82\\
         TargetVAE (ours) $\#$ 2 & 40.2 & 0.72 & 5.9 & 662 & 0.71 & 7.64 \\
         \bottomrule
    \end{tabular}
    \caption{Trade-off between binding affinities and pharmaceutical properties (i.e. QED and SA).}
    \label{tab:experimental_binding_2}
\end{table}

\section{Experiments on Atom3D Ligand Binding}
This section describes the experiments on the Ligand Binding Affinity (LBA) dataset in the Atom3D benchmark \cite{townshend2020atom3d}. In LBA, binding regions on the protein surface are determined and drug-target complexes are provided, and this alleviates the search for binding sites on the entire protein structure. Therefore, we relax the PMN model to only work on the 3D structure of protein-ligand complexes (i.e. the language modeling part is excluded) in which the input graphs are atom-level and constructed by combining the atoms of proteins and ligands. Edges are determined based on distances among them. We compare our model with existing methods, including ENN, 3D CNN, GNN, and GVP-MPNN. ENN refers to the Cormorant neural network \cite{NEURIPS2019_03573b32} that works on 3D point clouds via spherical harmonic convolutions. We train and evaluate our method according to the split-30-sequence settings provided in \cite{townshend2020atom3d} and compare the results with those reported in \cite{jing2021equivariant}. The models are evaluated by the root-mean-squared error (RMSE) metric. Moreover, 3D CNN and GNN are implemented in \cite{townshend2020atom3d}, and GVP-MPNN are proposed in \cite{jing2021equivariant}
On LBA, we use root-mean-squared error (RMSE) to compare ours with other baselines provided in \cite{townshend2020atom3d}. 

\subsection{Experimental Results} We run the method in four random seeds of weight initialization. Table \ref{results:lba_atom3d} shows the comparisons between our proposed model and the baselines. The numbers reveal that using Transformers to model the long-range interactions among the atoms helps our model remarkably outperform the local networks, especially GVP-MPNN, which shares the same input features as ours. Furthermore, using local message passing and efficient transformers (e.g., Performer \cite{choromanski2021rethinking}, Reformer \cite{Kitaev2020Reformer:}, etc) reduces the time complexity to linear with respect to the number of residues, which is more efficient than ENN and comparable to the others in terms of computations.

\begin{table}[h]
\small
  \caption{Experimental Results on LBA dataset in Atom3d. In this experiment, we detach the language modeling component for fair comparisons with the baselines, which are performed on 3D structures of protein-ligand complexes.} 
  \vspace{10pt}
  \label{results:lba_atom3d}
  \centering
  \begin{tabular}{cccccc}
    \toprule 
    Methods & 3D-CNN & GNN & ENN & GVP-MPNN & PMN$^*$ (ours)                  \\
    \cmidrule(r){1-6}
    RMSE $\downarrow$   & 1.416 $\pm$ 0.021 & 1.570 $\pm$ 0.025 &
1.568 $\pm$ 0.012 & 1.594 $\pm$ 0.073 & \textbf{1.353 $\pm$ 0.012} \\
    \bottomrule
  \end{tabular}
\end{table}

%% file: data_and_software.tex
\section{Data and Software Availability}

The data that we used for this study is publicly available at \url{https://github.com/vtarasv/3d-prot-dta} \cite{D3RA00281K} and \url{https://www.atom3d.ai/} \cite{townshend2020atom3d}. We release our data processing pipeline and software along with the installation instructions at \url{https://github.com/HySonLab/Ligand_Generation}. All experimental results and visualizations are reproducible given our software release.

%% file: paper.bbl
\providecommand{\latin}[1]{#1}
\makeatletter
\providecommand{\doi}
  {\begingroup\let\do\@makeother\dospecials
  \catcode`\{=1 \catcode`\}=2 \doi@aux}
\providecommand{\doi@aux}[1]{\endgroup\texttt{#1}}
\makeatother
\providecommand*\mcitethebibliography{\thebibliography}
\csname @ifundefined\endcsname{endmcitethebibliography}
  {\let\endmcitethebibliography\endthebibliography}{}
\begin{mcitethebibliography}{77}
\providecommand*\natexlab[1]{#1}
\providecommand*\mciteSetBstSublistMode[1]{}
\providecommand*\mciteSetBstMaxWidthForm[2]{}
\providecommand*\mciteBstWouldAddEndPuncttrue
  {\def\EndOfBibitem{\unskip.}}
\providecommand*\mciteBstWouldAddEndPunctfalse
  {\let\EndOfBibitem\relax}
\providecommand*\mciteSetBstMidEndSepPunct[3]{}
\providecommand*\mciteSetBstSublistLabelBeginEnd[3]{}
\providecommand*\EndOfBibitem{}
\mciteSetBstSublistMode{f}
\mciteSetBstMaxWidthForm{subitem}{(\alph{mcitesubitemcount})}
\mciteSetBstSublistLabelBeginEnd
  {\mcitemaxwidthsubitemform\space}
  {\relax}
  {\relax}

\bibitem[Hughes \latin{et~al.}(2011)Hughes, Rees, Kalindjian, and
  Philpott]{hughes2011principles}
Hughes,~J.~P.; Rees,~S.; Kalindjian,~S.~B.; Philpott,~K.~L. Principles of early
  drug discovery. \emph{British Journal of Pharmacology} \textbf{2011},
  \emph{162}, 1239--1249\relax
\mciteBstWouldAddEndPuncttrue
\mciteSetBstMidEndSepPunct{\mcitedefaultmidpunct}
{\mcitedefaultendpunct}{\mcitedefaultseppunct}\relax
\EndOfBibitem
\bibitem[Verkhivker \latin{et~al.}(2001)Verkhivker, Bouzida, Gehlhaar, Rejto,
  Arthurs, Colson, Freer, Larson, Luty, Marrone, \latin{et~al.}
  others]{verkhivker2001binding}
Verkhivker,~G.~M.; Bouzida,~D.; Gehlhaar,~D.~K.; Rejto,~P.~A.; Arthurs,~S.;
  Colson,~A.~B.; Freer,~S.~T.; Larson,~V.; Luty,~B.~A.; Marrone,~T.,
  \latin{et~al.}  \emph{Combinatorial Library Design and Evaluation}; CRC
  Press, 2001; pp 177--216\relax
\mciteBstWouldAddEndPuncttrue
\mciteSetBstMidEndSepPunct{\mcitedefaultmidpunct}
{\mcitedefaultendpunct}{\mcitedefaultseppunct}\relax
\EndOfBibitem
\bibitem[Burley \latin{et~al.}(2019)Burley, Berman, Bhikadiya, Bi, Chen,
  Di~Costanzo, Christie, Dalenberg, Duarte, Dutta, \latin{et~al.}
  others]{burley2019rcsb}
Burley,~S.~K.; Berman,~H.~M.; Bhikadiya,~C.; Bi,~C.; Chen,~L.; Di~Costanzo,~L.;
  Christie,~C.; Dalenberg,~K.; Duarte,~J.~M.; Dutta,~S., \latin{et~al.}  RCSB
  Protein Data Bank: biological macromolecular structures enabling research and
  education in fundamental biology, biomedicine, biotechnology and energy.
  \emph{Nucleic acids research} \textbf{2019}, \emph{47}, D464--D474\relax
\mciteBstWouldAddEndPuncttrue
\mciteSetBstMidEndSepPunct{\mcitedefaultmidpunct}
{\mcitedefaultendpunct}{\mcitedefaultseppunct}\relax
\EndOfBibitem
\bibitem[You \latin{et~al.}(2018)You, Liu, Ying, Pande, and
  Leskovec]{NEURIPS2018_d60678e8}
You,~J.; Liu,~B.; Ying,~Z.; Pande,~V.; Leskovec,~J. Graph Convolutional Policy
  Network for Goal-Directed Molecular Graph Generation. Advances in Neural
  Information Processing Systems. 2018\relax
\mciteBstWouldAddEndPuncttrue
\mciteSetBstMidEndSepPunct{\mcitedefaultmidpunct}
{\mcitedefaultendpunct}{\mcitedefaultseppunct}\relax
\EndOfBibitem
\bibitem[Jin \latin{et~al.}(2018)Jin, Barzilay, and Jaakkola]{pmlr-v80-jin18a}
Jin,~W.; Barzilay,~R.; Jaakkola,~T. Junction Tree Variational Autoencoder for
  Molecular Graph Generation. Proceedings of the 35th International Conference
  on Machine Learning. 2018; pp 2323--2332\relax
\mciteBstWouldAddEndPuncttrue
\mciteSetBstMidEndSepPunct{\mcitedefaultmidpunct}
{\mcitedefaultendpunct}{\mcitedefaultseppunct}\relax
\EndOfBibitem
\bibitem[Jin \latin{et~al.}(2020)Jin, Barzilay, and Jaakkola]{pmlr-v119-jin20a}
Jin,~W.; Barzilay,~D.; Jaakkola,~T. Hierarchical Generation of Molecular Graphs
  using Structural Motifs. Proceedings of the 37th International Conference on
  Machine Learning. 2020; pp 4839--4848\relax
\mciteBstWouldAddEndPuncttrue
\mciteSetBstMidEndSepPunct{\mcitedefaultmidpunct}
{\mcitedefaultendpunct}{\mcitedefaultseppunct}\relax
\EndOfBibitem
\bibitem[Luo \latin{et~al.}(2021)Luo, Guan, Ma, and Peng]{luo2021a}
Luo,~S.; Guan,~J.; Ma,~J.; Peng,~J. A 3D Generative Model for Structure-Based
  Drug Design. Advances in Neural Information Processing Systems. 2021\relax
\mciteBstWouldAddEndPuncttrue
\mciteSetBstMidEndSepPunct{\mcitedefaultmidpunct}
{\mcitedefaultendpunct}{\mcitedefaultseppunct}\relax
\EndOfBibitem
\bibitem[Simonovsky and Komodakis(2018)Simonovsky, and
  Komodakis]{Simonovsky2018GraphVAETG}
Simonovsky,~M.; Komodakis,~N. GraphVAE: Towards Generation of Small Graphs
  Using Variational Autoencoders. \emph{ArXiv} \textbf{2018},
  \emph{abs/1802.03480}\relax
\mciteBstWouldAddEndPuncttrue
\mciteSetBstMidEndSepPunct{\mcitedefaultmidpunct}
{\mcitedefaultendpunct}{\mcitedefaultseppunct}\relax
\EndOfBibitem
\bibitem[De~Cao and Kipf(2018)De~Cao, and Kipf]{de2018molgan}
De~Cao,~N.; Kipf,~T. {MolGAN: An implicit generative model for small molecular
  graphs}. \emph{ICML 2018 workshop on Theoretical Foundations and Applications
  of Deep Generative Models} \textbf{2018}, \relax
\mciteBstWouldAddEndPunctfalse
\mciteSetBstMidEndSepPunct{\mcitedefaultmidpunct}
{}{\mcitedefaultseppunct}\relax
\EndOfBibitem
\bibitem[Luo \latin{et~al.}(2021)Luo, Yan, and Ji]{pmlr-v139-luo21a}
Luo,~Y.; Yan,~K.; Ji,~S. GraphDF: A Discrete Flow Model for Molecular Graph
  Generation. Proceedings of the 38th International Conference on Machine
  Learning. 2021; pp 7192--7203\relax
\mciteBstWouldAddEndPuncttrue
\mciteSetBstMidEndSepPunct{\mcitedefaultmidpunct}
{\mcitedefaultendpunct}{\mcitedefaultseppunct}\relax
\EndOfBibitem
\bibitem[Gapsys \latin{et~al.}(2022)Gapsys, Hahn, Tresadern, Mobley, Rampp, and
  de~Groot]{gapsys2022pre}
Gapsys,~V.; Hahn,~D.~F.; Tresadern,~G.; Mobley,~D.~L.; Rampp,~M.;
  de~Groot,~B.~L. Pre-exascale computing of protein-ligand binding free
  energies with open source software for drug design. \emph{Journal of chemical
  information and modeling} \textbf{2022}, \emph{62}, 1172--1177\relax
\mciteBstWouldAddEndPuncttrue
\mciteSetBstMidEndSepPunct{\mcitedefaultmidpunct}
{\mcitedefaultendpunct}{\mcitedefaultseppunct}\relax
\EndOfBibitem
\bibitem[Berman \latin{et~al.}(2000)Berman, Westbrook, Feng, Gilliland, Bhat,
  Weissig, Shindyalov, and Bourne]{berman2000protein}
Berman,~H.~M.; Westbrook,~J.; Feng,~Z.; Gilliland,~G.; Bhat,~T.~N.;
  Weissig,~H.; Shindyalov,~I.~N.; Bourne,~P.~E. The protein data bank.
  \emph{Nucleic acids research} \textbf{2000}, \emph{28}, 235--242\relax
\mciteBstWouldAddEndPuncttrue
\mciteSetBstMidEndSepPunct{\mcitedefaultmidpunct}
{\mcitedefaultendpunct}{\mcitedefaultseppunct}\relax
\EndOfBibitem
\bibitem[Notin \latin{et~al.}(2022)Notin, Dias, Frazer, Hurtado, Gomez, Marks,
  and Gal]{pmlr-v162-notin22a}
Notin,~P.; Dias,~M.; Frazer,~J.; Hurtado,~J.~M.; Gomez,~A.~N.; Marks,~D.;
  Gal,~Y. Tranception: Protein Fitness Prediction with Autoregressive
  Transformers and Inference-time Retrieval. Proceedings of the 39th
  International Conference on Machine Learning. 2022; pp 16990--17017\relax
\mciteBstWouldAddEndPuncttrue
\mciteSetBstMidEndSepPunct{\mcitedefaultmidpunct}
{\mcitedefaultendpunct}{\mcitedefaultseppunct}\relax
\EndOfBibitem
\bibitem[Brandes \latin{et~al.}(2022)Brandes, Ofer, Peleg, Rappoport, and
  Linial]{10.1093/bioinformatics/btac020}
Brandes,~N.; Ofer,~D.; Peleg,~Y.; Rappoport,~N.; Linial,~M. {ProteinBERT: a
  universal deep-learning model of protein sequence and function}.
  \emph{Bioinformatics} \textbf{2022}, \emph{38}, 2102--2110\relax
\mciteBstWouldAddEndPuncttrue
\mciteSetBstMidEndSepPunct{\mcitedefaultmidpunct}
{\mcitedefaultendpunct}{\mcitedefaultseppunct}\relax
\EndOfBibitem
\bibitem[Asgari \latin{et~al.}(2019)Asgari, McHardy, and
  Mofrad]{asgari2019probabilistic}
Asgari,~E.; McHardy,~A.~C.; Mofrad,~M.~R. Probabilistic variable-length
  segmentation of protein sequences for discriminative motif discovery
  (DiMotif) and sequence embedding (ProtVecX). \emph{Scientific reports}
  \textbf{2019}, \emph{9}, 1--16\relax
\mciteBstWouldAddEndPuncttrue
\mciteSetBstMidEndSepPunct{\mcitedefaultmidpunct}
{\mcitedefaultendpunct}{\mcitedefaultseppunct}\relax
\EndOfBibitem
\bibitem[Wu \latin{et~al.}(2021)Wu, Johnston, Arnold, and Yang]{WU202118}
Wu,~Z.; Johnston,~K.~E.; Arnold,~F.~H.; Yang,~K.~K. Protein sequence design
  with deep generative models. \emph{Current Opinion in Chemical Biology}
  \textbf{2021}, \emph{65}, 18--27, Mechanistic Biology * Machine Learning in
  Chemical Biology\relax
\mciteBstWouldAddEndPuncttrue
\mciteSetBstMidEndSepPunct{\mcitedefaultmidpunct}
{\mcitedefaultendpunct}{\mcitedefaultseppunct}\relax
\EndOfBibitem
\bibitem[Yang \latin{et~al.}(2018)Yang, Wu, Bedbrook, and
  Arnold]{10.1093/bioinformatics/bty178}
Yang,~K.~K.; Wu,~Z.; Bedbrook,~C.~N.; Arnold,~F.~H. {Learned protein embeddings
  for machine learning}. \emph{Bioinformatics} \textbf{2018}, \emph{34},
  2642--2648\relax
\mciteBstWouldAddEndPuncttrue
\mciteSetBstMidEndSepPunct{\mcitedefaultmidpunct}
{\mcitedefaultendpunct}{\mcitedefaultseppunct}\relax
\EndOfBibitem
\bibitem[Anderson \latin{et~al.}(2019)Anderson, Hy, and
  Kondor]{NEURIPS2019_03573b32}
Anderson,~B.; Hy,~T.~S.; Kondor,~R. Cormorant: Covariant Molecular Neural
  Networks. Advances in Neural Information Processing Systems. 2019\relax
\mciteBstWouldAddEndPuncttrue
\mciteSetBstMidEndSepPunct{\mcitedefaultmidpunct}
{\mcitedefaultendpunct}{\mcitedefaultseppunct}\relax
\EndOfBibitem
\bibitem[Townshend \latin{et~al.}(2020)Townshend, V{\"o}gele, Suriana, Derry,
  Powers, Laloudakis, Balachandar, Jing, Anderson, Eismann, \latin{et~al.}
  others]{townshend2020atom3d}
Townshend,~R.~J.; V{\"o}gele,~M.; Suriana,~P.; Derry,~A.; Powers,~A.;
  Laloudakis,~Y.; Balachandar,~S.; Jing,~B.; Anderson,~B.; Eismann,~S.,
  \latin{et~al.}  Atom3d: Tasks on molecules in three dimensions. \emph{arXiv
  preprint arXiv:2012.04035} \textbf{2020}, \relax
\mciteBstWouldAddEndPunctfalse
\mciteSetBstMidEndSepPunct{\mcitedefaultmidpunct}
{}{\mcitedefaultseppunct}\relax
\EndOfBibitem
\bibitem[Jing \latin{et~al.}(2021)Jing, Eismann, Soni, and
  Dror]{jing2021equivariant}
Jing,~B.; Eismann,~S.; Soni,~P.~N.; Dror,~R.~O. Equivariant graph neural
  networks for 3d macromolecular structure. \emph{arXiv preprint
  arXiv:2106.03843} \textbf{2021}, \relax
\mciteBstWouldAddEndPunctfalse
\mciteSetBstMidEndSepPunct{\mcitedefaultmidpunct}
{}{\mcitedefaultseppunct}\relax
\EndOfBibitem
\bibitem[Jing \latin{et~al.}(2021)Jing, Eismann, Suriana, Townshend, and
  Dror]{jing2021learning}
Jing,~B.; Eismann,~S.; Suriana,~P.; Townshend,~R. J.~L.; Dror,~R. Learning from
  Protein Structure with Geometric Vector Perceptrons. International Conference
  on Learning Representations. 2021\relax
\mciteBstWouldAddEndPuncttrue
\mciteSetBstMidEndSepPunct{\mcitedefaultmidpunct}
{\mcitedefaultendpunct}{\mcitedefaultseppunct}\relax
\EndOfBibitem
\bibitem[Zhao \latin{et~al.}(2022)Zhao, Liu, and Wang]{10.1093/nargab/lqac004}
Zhao,~C.; Liu,~T.; Wang,~Z. {PANDA2: protein function prediction using graph
  neural networks}. \emph{NAR Genomics and Bioinformatics} \textbf{2022},
  \emph{4}, lqac004\relax
\mciteBstWouldAddEndPuncttrue
\mciteSetBstMidEndSepPunct{\mcitedefaultmidpunct}
{\mcitedefaultendpunct}{\mcitedefaultseppunct}\relax
\EndOfBibitem
\bibitem[Guan \latin{et~al.}(2023)Guan, Qian, Peng, Su, Peng, and Ma]{guan3d}
Guan,~J.; Qian,~W.~W.; Peng,~X.; Su,~Y.; Peng,~J.; Ma,~J. 3D Equivariant
  Diffusion for Target-Aware Molecule Generation and Affinity Prediction.
  International Conference on Learning Representations. 2023\relax
\mciteBstWouldAddEndPuncttrue
\mciteSetBstMidEndSepPunct{\mcitedefaultmidpunct}
{\mcitedefaultendpunct}{\mcitedefaultseppunct}\relax
\EndOfBibitem
\bibitem[Peng \latin{et~al.}(2022)Peng, Luo, Guan, Xie, Peng, and
  Ma]{pmlr-v162-peng22b}
Peng,~X.; Luo,~S.; Guan,~J.; Xie,~Q.; Peng,~J.; Ma,~J. {P}ocket2{M}ol:
  Efficient Molecular Sampling Based on 3{D} Protein Pockets. Proceedings of
  the 39th International Conference on Machine Learning. 2022; pp
  17644--17655\relax
\mciteBstWouldAddEndPuncttrue
\mciteSetBstMidEndSepPunct{\mcitedefaultmidpunct}
{\mcitedefaultendpunct}{\mcitedefaultseppunct}\relax
\EndOfBibitem
\bibitem[Luo \latin{et~al.}(2021)Luo, Guan, Ma, and Peng]{NEURIPS2021_31445061}
Luo,~S.; Guan,~J.; Ma,~J.; Peng,~J. A 3D Generative Model for Structure-Based
  Drug Design. Advances in Neural Information Processing Systems. 2021; pp
  6229--6239\relax
\mciteBstWouldAddEndPuncttrue
\mciteSetBstMidEndSepPunct{\mcitedefaultmidpunct}
{\mcitedefaultendpunct}{\mcitedefaultseppunct}\relax
\EndOfBibitem
\bibitem[Liu \latin{et~al.}(2022)Liu, Luo, Uchino, Maruhashi, and
  Ji]{pmlr-v162-liu22m}
Liu,~M.; Luo,~Y.; Uchino,~K.; Maruhashi,~K.; Ji,~S. Generating 3{D} Molecules
  for Target Protein Binding. Proceedings of the 39th International Conference
  on Machine Learning. 2022; pp 13912--13924\relax
\mciteBstWouldAddEndPuncttrue
\mciteSetBstMidEndSepPunct{\mcitedefaultmidpunct}
{\mcitedefaultendpunct}{\mcitedefaultseppunct}\relax
\EndOfBibitem
\bibitem[Scantlebury \latin{et~al.}(2023)Scantlebury, Vost, Carbery, Hadfield,
  Turnbull, Brown, Chenthamarakshan, Das, Grosjean, von Delft, and
  Deane]{Scantlebury2023}
Scantlebury,~J.; Vost,~L.; Carbery,~A.; Hadfield,~T.~E.; Turnbull,~O.~M.;
  Brown,~N.; Chenthamarakshan,~V.; Das,~P.; Grosjean,~H.; von Delft,~F.;
  Deane,~C.~M. A Small Step Toward Generalizability: Training a Machine
  Learning Scoring Function for Structure-Based Virtual Screening.
  \emph{Journal of Chemical Information and Modeling} \textbf{2023}, \emph{63},
  2960--2974\relax
\mciteBstWouldAddEndPuncttrue
\mciteSetBstMidEndSepPunct{\mcitedefaultmidpunct}
{\mcitedefaultendpunct}{\mcitedefaultseppunct}\relax
\EndOfBibitem
\bibitem[Nascimento \latin{et~al.}(2016)Nascimento, Prud{\^e}ncio, and
  Costa]{nascimento2016multiple}
Nascimento,~A.~C.; Prud{\^e}ncio,~R.~B.; Costa,~I.~G. A multiple kernel
  learning algorithm for drug-target interaction prediction. \emph{BMC
  bioinformatics} \textbf{2016}, \emph{17}, 1--16\relax
\mciteBstWouldAddEndPuncttrue
\mciteSetBstMidEndSepPunct{\mcitedefaultmidpunct}
{\mcitedefaultendpunct}{\mcitedefaultseppunct}\relax
\EndOfBibitem
\bibitem[He \latin{et~al.}(2017)He, Heidemeyer, Ban, Cherkasov, and
  Ester]{He2017}
He,~T.; Heidemeyer,~M.; Ban,~F.; Cherkasov,~A.; Ester,~M. SimBoost: a
  read-across approach for predicting drug--target binding affinities using
  gradient boosting machines. \emph{Journal of Cheminformatics} \textbf{2017},
  \emph{9}, 24\relax
\mciteBstWouldAddEndPuncttrue
\mciteSetBstMidEndSepPunct{\mcitedefaultmidpunct}
{\mcitedefaultendpunct}{\mcitedefaultseppunct}\relax
\EndOfBibitem
\bibitem[Öztürk \latin{et~al.}(2018)Öztürk, Özgür, and
  Ozkirimli]{10.1093/bioinformatics/bty593}
Öztürk,~H.; Özgür,~A.; Ozkirimli,~E. {DeepDTA: deep drug–target binding
  affinity prediction}. \emph{Bioinformatics} \textbf{2018}, \emph{34},
  i821--i829\relax
\mciteBstWouldAddEndPuncttrue
\mciteSetBstMidEndSepPunct{\mcitedefaultmidpunct}
{\mcitedefaultendpunct}{\mcitedefaultseppunct}\relax
\EndOfBibitem
\bibitem[Zhao \latin{et~al.}(2020)Zhao, Wang, Pang, Liu, and
  Zhang]{Zhao2020-gx}
Zhao,~L.; Wang,~J.; Pang,~L.; Liu,~Y.; Zhang,~J. {GANsDTA}: Predicting
  {Drug-Target} Binding Affinity Using {GANs}. \emph{Front Genet}
  \textbf{2020}, \emph{10}, 1243\relax
\mciteBstWouldAddEndPuncttrue
\mciteSetBstMidEndSepPunct{\mcitedefaultmidpunct}
{\mcitedefaultendpunct}{\mcitedefaultseppunct}\relax
\EndOfBibitem
\bibitem[Nguyen \latin{et~al.}(2020)Nguyen, Le, Quinn, Nguyen, Le, and
  Venkatesh]{10.1093/bioinformatics/btaa921}
Nguyen,~T.; Le,~H.; Quinn,~T.~P.; Nguyen,~T.; Le,~T.~D.; Venkatesh,~S.
  {GraphDTA: predicting drug–target binding affinity with graph neural
  networks}. \emph{Bioinformatics} \textbf{2020}, \emph{37}, 1140--1147\relax
\mciteBstWouldAddEndPuncttrue
\mciteSetBstMidEndSepPunct{\mcitedefaultmidpunct}
{\mcitedefaultendpunct}{\mcitedefaultseppunct}\relax
\EndOfBibitem
\bibitem[Voitsitskyi \latin{et~al.}(2023)Voitsitskyi, Stratiichuk, Koleiev,
  Popryho, Ostrovsky, Henitsoi, Khropachov, Vozniak, Zhytar, Nechepurenko,
  Yesylevskyy, Nafiiev, and Starosyla]{D3RA00281K}
Voitsitskyi,~T.; Stratiichuk,~R.; Koleiev,~I.; Popryho,~L.; Ostrovsky,~Z.;
  Henitsoi,~P.; Khropachov,~I.; Vozniak,~V.; Zhytar,~R.; Nechepurenko,~D.;
  Yesylevskyy,~S.; Nafiiev,~A.; Starosyla,~S. 3DProtDTA: a deep learning model
  for drug-target affinity prediction based on residue-level protein graphs.
  \emph{RSC Adv.} \textbf{2023}, \emph{13}, 10261--10272\relax
\mciteBstWouldAddEndPuncttrue
\mciteSetBstMidEndSepPunct{\mcitedefaultmidpunct}
{\mcitedefaultendpunct}{\mcitedefaultseppunct}\relax
\EndOfBibitem
\bibitem[Merz~Jr. \latin{et~al.}(2020)Merz~Jr., De~Fabritiis, and
  Wei]{Merz2020}
Merz~Jr.,~K.~M.; De~Fabritiis,~G.; Wei,~G.-W. Generative Models for Molecular
  Design. \emph{Journal of Chemical Information and Modeling} \textbf{2020},
  \emph{60}, 5635--5636\relax
\mciteBstWouldAddEndPuncttrue
\mciteSetBstMidEndSepPunct{\mcitedefaultmidpunct}
{\mcitedefaultendpunct}{\mcitedefaultseppunct}\relax
\EndOfBibitem
\bibitem[Gómez-Bombarelli \latin{et~al.}(2018)Gómez-Bombarelli, Wei,
  Duvenaud, Hernández-Lobato, Sánchez-Lengeling, Sheberla,
  Aguilera-Iparraguirre, Hirzel, Adams, and
  Aspuru-Guzik]{doi:10.1021/acscentsci.7b00572}
Gómez-Bombarelli,~R.; Wei,~J.~N.; Duvenaud,~D.; Hernández-Lobato,~J.~M.;
  Sánchez-Lengeling,~B.; Sheberla,~D.; Aguilera-Iparraguirre,~J.;
  Hirzel,~T.~D.; Adams,~R.~P.; Aspuru-Guzik,~A. Automatic Chemical Design Using
  a Data-Driven Continuous Representation of Molecules. \emph{ACS Central
  Science} \textbf{2018}, \emph{4}, 268--276, PMID: 29532027\relax
\mciteBstWouldAddEndPuncttrue
\mciteSetBstMidEndSepPunct{\mcitedefaultmidpunct}
{\mcitedefaultendpunct}{\mcitedefaultseppunct}\relax
\EndOfBibitem
\bibitem[Segler \latin{et~al.}(2018)Segler, Kogej, Tyrchan, and
  Waller]{doi:10.1021/acscentsci.7b00512}
Segler,~M. H.~S.; Kogej,~T.; Tyrchan,~C.; Waller,~M.~P. Generating Focused
  Molecule Libraries for Drug Discovery with Recurrent Neural Networks.
  \emph{ACS Central Science} \textbf{2018}, \emph{4}, 120--131, PMID:
  29392184\relax
\mciteBstWouldAddEndPuncttrue
\mciteSetBstMidEndSepPunct{\mcitedefaultmidpunct}
{\mcitedefaultendpunct}{\mcitedefaultseppunct}\relax
\EndOfBibitem
\bibitem[Gao \latin{et~al.}(2020)Gao, Nguyen, Tu, and Wei]{Gao2020}
Gao,~K.; Nguyen,~D.~D.; Tu,~M.; Wei,~G.-W. Generative Network Complex for the
  Automated Generation of Drug-like Molecules. \emph{Journal of Chemical
  Information and Modeling} \textbf{2020}, \emph{60}, 5682--5698\relax
\mciteBstWouldAddEndPuncttrue
\mciteSetBstMidEndSepPunct{\mcitedefaultmidpunct}
{\mcitedefaultendpunct}{\mcitedefaultseppunct}\relax
\EndOfBibitem
\bibitem[Kusner \latin{et~al.}(2017)Kusner, Paige, and
  Hern{\'a}ndez-Lobato]{pmlr-v70-kusner17a}
Kusner,~M.~J.; Paige,~B.; Hern{\'a}ndez-Lobato,~J.~M. Grammar Variational
  Autoencoder. Proceedings of the 34th International Conference on Machine
  Learning. 2017; pp 1945--1954\relax
\mciteBstWouldAddEndPuncttrue
\mciteSetBstMidEndSepPunct{\mcitedefaultmidpunct}
{\mcitedefaultendpunct}{\mcitedefaultseppunct}\relax
\EndOfBibitem
\bibitem[Dai \latin{et~al.}(2018)Dai, Tian, Dai, Skiena, and
  Song]{dai2018syntaxdirected}
Dai,~H.; Tian,~Y.; Dai,~B.; Skiena,~S.; Song,~L. Syntax-Directed Variational
  Autoencoder for Structured Data. International Conference on Learning
  Representations. 2018\relax
\mciteBstWouldAddEndPuncttrue
\mciteSetBstMidEndSepPunct{\mcitedefaultmidpunct}
{\mcitedefaultendpunct}{\mcitedefaultseppunct}\relax
\EndOfBibitem
\bibitem[Thiede \latin{et~al.}(2020)Thiede, Hy, and Kondor]{thiede2020general}
Thiede,~E.~H.; Hy,~T.~S.; Kondor,~R. The general theory of permutation
  equivarant neural networks and higher order graph variational encoders.
  \emph{arXiv preprint arXiv:2004.03990} \textbf{2020}, \relax
\mciteBstWouldAddEndPunctfalse
\mciteSetBstMidEndSepPunct{\mcitedefaultmidpunct}
{}{\mcitedefaultseppunct}\relax
\EndOfBibitem
\bibitem[Hy and Kondor(2023)Hy, and Kondor]{Hy_2023}
Hy,~T.~S.; Kondor,~R. Multiresolution equivariant graph variational
  autoencoder. \emph{Machine Learning: Science and Technology} \textbf{2023},
  \emph{4}, 015031\relax
\mciteBstWouldAddEndPuncttrue
\mciteSetBstMidEndSepPunct{\mcitedefaultmidpunct}
{\mcitedefaultendpunct}{\mcitedefaultseppunct}\relax
\EndOfBibitem
\bibitem[Krenn \latin{et~al.}(2020)Krenn, Häse, Nigam, Friederich, and
  Aspuru-Guzik]{Krenn_2020}
Krenn,~M.; Häse,~F.; Nigam,~A.; Friederich,~P.; Aspuru-Guzik,~A.
  Self-referencing embedded strings (SELFIES): A 100$\%$ robust molecular
  string representation. \emph{Machine Learning: Science and Technology}
  \textbf{2020}, \emph{1}, 045024\relax
\mciteBstWouldAddEndPuncttrue
\mciteSetBstMidEndSepPunct{\mcitedefaultmidpunct}
{\mcitedefaultendpunct}{\mcitedefaultseppunct}\relax
\EndOfBibitem
\bibitem[Harvey \latin{et~al.}(2022)Harvey, Naderiparizi, and
  Wood]{harvey2022conditional}
Harvey,~W.; Naderiparizi,~S.; Wood,~F. Conditional Image Generation by
  Conditioning Variational Auto-Encoders. International Conference on Learning
  Representations. 2022\relax
\mciteBstWouldAddEndPuncttrue
\mciteSetBstMidEndSepPunct{\mcitedefaultmidpunct}
{\mcitedefaultendpunct}{\mcitedefaultseppunct}\relax
\EndOfBibitem
\bibitem[Kingma and Welling(2013)Kingma, and Welling]{kingma2013auto}
Kingma,~D.~P.; Welling,~M. Auto-encoding variational bayes. \emph{arXiv
  preprint arXiv:1312.6114} \textbf{2013}, \relax
\mciteBstWouldAddEndPunctfalse
\mciteSetBstMidEndSepPunct{\mcitedefaultmidpunct}
{}{\mcitedefaultseppunct}\relax
\EndOfBibitem
\bibitem[Sohn \latin{et~al.}(2015)Sohn, Lee, and Yan]{NIPS2015_8d55a249}
Sohn,~K.; Lee,~H.; Yan,~X. Learning Structured Output Representation using Deep
  Conditional Generative Models. Advances in Neural Information Processing
  Systems. 2015\relax
\mciteBstWouldAddEndPuncttrue
\mciteSetBstMidEndSepPunct{\mcitedefaultmidpunct}
{\mcitedefaultendpunct}{\mcitedefaultseppunct}\relax
\EndOfBibitem
\bibitem[Zheng \latin{et~al.}(2019)Zheng, Cham, and Cai]{Zheng_2019_CVPR}
Zheng,~C.; Cham,~T.-J.; Cai,~J. Pluralistic Image Completion. Proceedings of
  the IEEE/CVF Conference on Computer Vision and Pattern Recognition (CVPR).
  2019\relax
\mciteBstWouldAddEndPuncttrue
\mciteSetBstMidEndSepPunct{\mcitedefaultmidpunct}
{\mcitedefaultendpunct}{\mcitedefaultseppunct}\relax
\EndOfBibitem
\bibitem[Ivanov \latin{et~al.}(2019)Ivanov, Figurnov, and
  Vetrov]{ivanov2018variational}
Ivanov,~O.; Figurnov,~M.; Vetrov,~D. Variational Autoencoder with Arbitrary
  Conditioning. International Conference on Learning Representations.
  2019\relax
\mciteBstWouldAddEndPuncttrue
\mciteSetBstMidEndSepPunct{\mcitedefaultmidpunct}
{\mcitedefaultendpunct}{\mcitedefaultseppunct}\relax
\EndOfBibitem
\bibitem[Wan \latin{et~al.}(2021)Wan, Zhang, Chen, and Liao]{wan2021high}
Wan,~Z.; Zhang,~J.; Chen,~D.; Liao,~J. High-Fidelity Pluralistic Image
  Completion with Transformers. \emph{arXiv preprint arXiv:2103.14031}
  \textbf{2021}, \relax
\mciteBstWouldAddEndPunctfalse
\mciteSetBstMidEndSepPunct{\mcitedefaultmidpunct}
{}{\mcitedefaultseppunct}\relax
\EndOfBibitem
\bibitem[Gilmer \latin{et~al.}(2017)Gilmer, Schoenholz, Riley, Vinyals, and
  Dahl]{10.5555/3305381.3305512}
Gilmer,~J.; Schoenholz,~S.~S.; Riley,~P.~F.; Vinyals,~O.; Dahl,~G.~E. Neural
  Message Passing for Quantum Chemistry. Proceedings of the 34th International
  Conference on Machine Learning - Volume 70. 2017; p 1263–1272\relax
\mciteBstWouldAddEndPuncttrue
\mciteSetBstMidEndSepPunct{\mcitedefaultmidpunct}
{\mcitedefaultendpunct}{\mcitedefaultseppunct}\relax
\EndOfBibitem
\bibitem[Dwivedi \latin{et~al.}(2022)Dwivedi, Ramp\'{a}\v{s}ek, Galkin, Parviz,
  Wolf, Luu, and Beaini]{NEURIPS2022_8c3c6668}
Dwivedi,~V.~P.; Ramp\'{a}\v{s}ek,~L.; Galkin,~M.; Parviz,~A.; Wolf,~G.;
  Luu,~A.~T.; Beaini,~D. Long Range Graph Benchmark. Advances in Neural
  Information Processing Systems. 2022; pp 22326--22340\relax
\mciteBstWouldAddEndPuncttrue
\mciteSetBstMidEndSepPunct{\mcitedefaultmidpunct}
{\mcitedefaultendpunct}{\mcitedefaultseppunct}\relax
\EndOfBibitem
\bibitem[Ngo \latin{et~al.}(2023)Ngo, Hy, and Kondor]{10.1063/5.0152833}
Ngo,~N.~K.; Hy,~T.~S.; Kondor,~R. {Multiresolution graph transformers and
  wavelet positional encoding for learning long-range and hierarchical
  structures}. \emph{The Journal of Chemical Physics} \textbf{2023},
  \emph{159}, 034109\relax
\mciteBstWouldAddEndPuncttrue
\mciteSetBstMidEndSepPunct{\mcitedefaultmidpunct}
{\mcitedefaultendpunct}{\mcitedefaultseppunct}\relax
\EndOfBibitem
\bibitem[Chen \latin{et~al.}(2020)Chen, Lin, Li, Li, Zhou, and
  Sun]{Chen_Lin_Li_Li_Zhou_Sun_2020}
Chen,~D.; Lin,~Y.; Li,~W.; Li,~P.; Zhou,~J.; Sun,~X. Measuring and Relieving
  the Over-Smoothing Problem for Graph Neural Networks from the Topological
  View. \emph{Proceedings of the AAAI Conference on Artificial Intelligence}
  \textbf{2020}, \emph{34}, 3438--3445\relax
\mciteBstWouldAddEndPuncttrue
\mciteSetBstMidEndSepPunct{\mcitedefaultmidpunct}
{\mcitedefaultendpunct}{\mcitedefaultseppunct}\relax
\EndOfBibitem
\bibitem[Topping \latin{et~al.}(2022)Topping, Giovanni, Chamberlain, Dong, and
  Bronstein]{topping2022understanding}
Topping,~J.; Giovanni,~F.~D.; Chamberlain,~B.~P.; Dong,~X.; Bronstein,~M.~M.
  Understanding over-squashing and bottlenecks on graphs via curvature.
  International Conference on Learning Representations. 2022\relax
\mciteBstWouldAddEndPuncttrue
\mciteSetBstMidEndSepPunct{\mcitedefaultmidpunct}
{\mcitedefaultendpunct}{\mcitedefaultseppunct}\relax
\EndOfBibitem
\bibitem[Kim \latin{et~al.}(2022)Kim, Nguyen, Min, Cho, Lee, Lee, and
  Hong]{kim2022pure}
Kim,~J.; Nguyen,~D.~T.; Min,~S.; Cho,~S.; Lee,~M.; Lee,~H.; Hong,~S. Pure
  Transformers are Powerful Graph Learners. Advances in Neural Information
  Processing Systems. 2022\relax
\mciteBstWouldAddEndPuncttrue
\mciteSetBstMidEndSepPunct{\mcitedefaultmidpunct}
{\mcitedefaultendpunct}{\mcitedefaultseppunct}\relax
\EndOfBibitem
\bibitem[Cai \latin{et~al.}(2023)Cai, Hy, Yu, and Wang]{cai2023connection}
Cai,~C.; Hy,~T.~S.; Yu,~R.; Wang,~Y. On the Connection Between MPNN and Graph
  Transformer. \emph{International Conference of Machine Learning}
  \textbf{2023}, \relax
\mciteBstWouldAddEndPunctfalse
\mciteSetBstMidEndSepPunct{\mcitedefaultmidpunct}
{}{\mcitedefaultseppunct}\relax
\EndOfBibitem
\bibitem[Roy* \latin{et~al.}(2020)Roy*, Saffar*, Grangier, and
  Vaswani]{roy*2020efficient}
Roy*,~A.; Saffar*,~M.~T.; Grangier,~D.; Vaswani,~A. Efficient Content-Based
  Sparse Attention with Routing Transformers. 2020;
  \url{https://openreview.net/forum?id=B1gjs6EtDr}\relax
\mciteBstWouldAddEndPuncttrue
\mciteSetBstMidEndSepPunct{\mcitedefaultmidpunct}
{\mcitedefaultendpunct}{\mcitedefaultseppunct}\relax
\EndOfBibitem
\bibitem[Choromanski \latin{et~al.}(2021)Choromanski, Likhosherstov, Dohan,
  Song, Gane, Sarlos, Hawkins, Davis, Mohiuddin, Kaiser, Belanger, Colwell, and
  Weller]{choromanski2021rethinking}
Choromanski,~K.~M.; Likhosherstov,~V.; Dohan,~D.; Song,~X.; Gane,~A.;
  Sarlos,~T.; Hawkins,~P.; Davis,~J.~Q.; Mohiuddin,~A.; Kaiser,~L.;
  Belanger,~D.~B.; Colwell,~L.~J.; Weller,~A. Rethinking Attention with
  Performers. International Conference on Learning Representations. 2021\relax
\mciteBstWouldAddEndPuncttrue
\mciteSetBstMidEndSepPunct{\mcitedefaultmidpunct}
{\mcitedefaultendpunct}{\mcitedefaultseppunct}\relax
\EndOfBibitem
\bibitem[Kitaev \latin{et~al.}(2020)Kitaev, Kaiser, and
  Levskaya]{Kitaev2020Reformer:}
Kitaev,~N.; Kaiser,~L.; Levskaya,~A. Reformer: The Efficient Transformer.
  International Conference on Learning Representations. 2020\relax
\mciteBstWouldAddEndPuncttrue
\mciteSetBstMidEndSepPunct{\mcitedefaultmidpunct}
{\mcitedefaultendpunct}{\mcitedefaultseppunct}\relax
\EndOfBibitem
\bibitem[Rogers and Hahn(2010)Rogers, and Hahn]{rogers2010extended}
Rogers,~D.; Hahn,~M. Extended-connectivity fingerprints. \emph{Journal of
  chemical information and modeling} \textbf{2010}, \emph{50}, 742--754\relax
\mciteBstWouldAddEndPuncttrue
\mciteSetBstMidEndSepPunct{\mcitedefaultmidpunct}
{\mcitedefaultendpunct}{\mcitedefaultseppunct}\relax
\EndOfBibitem
\bibitem[Paszke \latin{et~al.}(2019)Paszke, Gross, Massa, Lerer, Bradbury,
  Chanan, Killeen, Lin, Gimelshein, Antiga, Desmaison, Kopf, Yang, DeVito,
  Raison, Tejani, Chilamkurthy, Steiner, Fang, Bai, and
  Chintala]{NEURIPS2019_bdbca288}
Paszke,~A. \latin{et~al.}  PyTorch: An Imperative Style, High-Performance Deep
  Learning Library. Advances in Neural Information Processing Systems.
  2019\relax
\mciteBstWouldAddEndPuncttrue
\mciteSetBstMidEndSepPunct{\mcitedefaultmidpunct}
{\mcitedefaultendpunct}{\mcitedefaultseppunct}\relax
\EndOfBibitem
\bibitem[Fey and Lenssen(2019)Fey, and Lenssen]{Fey:2019wv}
Fey,~M.; Lenssen,~J.~E. Fast Graph Representation Learning with {PyTorch
  Geometric}. ICLR 2019 Workshop on Representation Learning on Graphs and
  Manifolds. 2019\relax
\mciteBstWouldAddEndPuncttrue
\mciteSetBstMidEndSepPunct{\mcitedefaultmidpunct}
{\mcitedefaultendpunct}{\mcitedefaultseppunct}\relax
\EndOfBibitem
\bibitem[Davis \latin{et~al.}(2011)Davis, Hunt, Herrgard, Ciceri, Wodicka,
  Pallares, Hocker, Treiber, and Zarrinkar]{Davis2011-fb}
Davis,~M.~I.; Hunt,~J.~P.; Herrgard,~S.; Ciceri,~P.; Wodicka,~L.~M.;
  Pallares,~G.; Hocker,~M.; Treiber,~D.~K.; Zarrinkar,~P.~P. Comprehensive
  analysis of kinase inhibitor selectivity. \emph{Nature Biotechnology}
  \textbf{2011}, \emph{29}, 1046--1051\relax
\mciteBstWouldAddEndPuncttrue
\mciteSetBstMidEndSepPunct{\mcitedefaultmidpunct}
{\mcitedefaultendpunct}{\mcitedefaultseppunct}\relax
\EndOfBibitem
\bibitem[Tang \latin{et~al.}(2014)Tang, Szwajda, Shakyawar, Xu, Hintsanen,
  Wennerberg, and Aittokallio]{tang2014making}
Tang,~J.; Szwajda,~A.; Shakyawar,~S.; Xu,~T.; Hintsanen,~P.; Wennerberg,~K.;
  Aittokallio,~T. Making sense of large-scale kinase inhibitor bioactivity data
  sets: a comparative and integrative analysis. \emph{Journal of Chemical
  Information and Modeling} \textbf{2014}, \emph{54}, 735--743\relax
\mciteBstWouldAddEndPuncttrue
\mciteSetBstMidEndSepPunct{\mcitedefaultmidpunct}
{\mcitedefaultendpunct}{\mcitedefaultseppunct}\relax
\EndOfBibitem
\bibitem[{\"O}zt{\"u}rk \latin{et~al.}(2019){\"O}zt{\"u}rk, Ozkirimli, and
  {\"O}zg{\"u}r]{ozturk2019widedta}
{\"O}zt{\"u}rk,~H.; Ozkirimli,~E.; {\"O}zg{\"u}r,~A. WideDTA: prediction of
  drug-target binding affinity. \emph{arXiv preprint arXiv:1902.04166}
  \textbf{2019}, \relax
\mciteBstWouldAddEndPunctfalse
\mciteSetBstMidEndSepPunct{\mcitedefaultmidpunct}
{}{\mcitedefaultseppunct}\relax
\EndOfBibitem
\bibitem[Kalemati \latin{et~al.}(2023)Kalemati, Zamani~Emani, and
  Koohi]{10.1371/journal.pcbi.1011036}
Kalemati,~M.; Zamani~Emani,~M.; Koohi,~S. BiComp-DTA: Drug-target binding
  affinity prediction through complementary biological-related and
  compression-based featurization approach. \emph{PLOS Computational Biology}
  \textbf{2023}, \emph{19}, 1--28\relax
\mciteBstWouldAddEndPuncttrue
\mciteSetBstMidEndSepPunct{\mcitedefaultmidpunct}
{\mcitedefaultendpunct}{\mcitedefaultseppunct}\relax
\EndOfBibitem
\bibitem[Rives \latin{et~al.}(2021)Rives, Meier, Sercu, Goyal, Lin, Liu, Guo,
  Ott, Zitnick, Ma, \latin{et~al.} others]{rives2021biological}
Rives,~A.; Meier,~J.; Sercu,~T.; Goyal,~S.; Lin,~Z.; Liu,~J.; Guo,~D.; Ott,~M.;
  Zitnick,~C.~L.; Ma,~J., \latin{et~al.}  Biological structure and function
  emerge from scaling unsupervised learning to 250 million protein sequences.
  \emph{Proceedings of the National Academy of Sciences} \textbf{2021},
  \emph{118}, e2016239118\relax
\mciteBstWouldAddEndPuncttrue
\mciteSetBstMidEndSepPunct{\mcitedefaultmidpunct}
{\mcitedefaultendpunct}{\mcitedefaultseppunct}\relax
\EndOfBibitem
\bibitem[Shim \latin{et~al.}(2021)Shim, Hong, Sohn, and
  Hwang]{shim2021prediction}
Shim,~J.; Hong,~Z.-Y.; Sohn,~I.; Hwang,~C. Prediction of drug--target binding
  affinity using similarity-based convolutional neural network.
  \emph{Scientific Reports} \textbf{2021}, \emph{11}, 4416\relax
\mciteBstWouldAddEndPuncttrue
\mciteSetBstMidEndSepPunct{\mcitedefaultmidpunct}
{\mcitedefaultendpunct}{\mcitedefaultseppunct}\relax
\EndOfBibitem
\bibitem[Zhao \latin{et~al.}(2019)Zhao, Xiao, Yang, Li, and
  Wang]{zhao2019attentiondta}
Zhao,~Q.; Xiao,~F.; Yang,~M.; Li,~Y.; Wang,~J. AttentionDTA: prediction of
  drug--target binding affinity using attention model. 2019 IEEE international
  conference on Bioinformatics and Biomedicine (BIBM). 2019; pp 64--69\relax
\mciteBstWouldAddEndPuncttrue
\mciteSetBstMidEndSepPunct{\mcitedefaultmidpunct}
{\mcitedefaultendpunct}{\mcitedefaultseppunct}\relax
\EndOfBibitem
\bibitem[Zeng \latin{et~al.}(2021)Zeng, Chen, Luo, Li, and
  Peng]{10.1093/bib/bbab117}
Zeng,~Y.; Chen,~X.; Luo,~Y.; Li,~X.; Peng,~D. {Deep drug-target binding
  affinity prediction with multiple attention blocks}. \emph{Briefings in
  Bioinformatics} \textbf{2021}, \emph{22}, bbab117\relax
\mciteBstWouldAddEndPuncttrue
\mciteSetBstMidEndSepPunct{\mcitedefaultmidpunct}
{\mcitedefaultendpunct}{\mcitedefaultseppunct}\relax
\EndOfBibitem
\bibitem[Yuan \latin{et~al.}(2021)Yuan, Chen, and Chen]{10.1093/bib/bbab506}
Yuan,~W.; Chen,~G.; Chen,~C. Y.-C. {FusionDTA: attention-based feature
  polymerizer and knowledge distillation for drug-target binding affinity
  prediction}. \emph{Briefings in Bioinformatics} \textbf{2021}, \emph{23},
  bbab506\relax
\mciteBstWouldAddEndPuncttrue
\mciteSetBstMidEndSepPunct{\mcitedefaultmidpunct}
{\mcitedefaultendpunct}{\mcitedefaultseppunct}\relax
\EndOfBibitem
\bibitem[Preuer \latin{et~al.}(2018)Preuer, Renz, Unterthiner, Hochreiter, and
  Klambauer]{Preuer2018}
Preuer,~K.; Renz,~P.; Unterthiner,~T.; Hochreiter,~S.; Klambauer,~G.
  Fr{\'e}chet ChemNet Distance: A Metric for Generative Models for Molecules in
  Drug Discovery. \emph{Journal of Chemical Information and Modeling}
  \textbf{2018}, \emph{58}, 1736--1741\relax
\mciteBstWouldAddEndPuncttrue
\mciteSetBstMidEndSepPunct{\mcitedefaultmidpunct}
{\mcitedefaultendpunct}{\mcitedefaultseppunct}\relax
\EndOfBibitem
\bibitem[Eckmann \latin{et~al.}(2022)Eckmann, Sun, Zhao, Feng, Gilson, and
  Yu]{eckmann2022limo}
Eckmann,~P.; Sun,~K.; Zhao,~B.; Feng,~M.; Gilson,~M.~K.; Yu,~R. LIMO: Latent
  Inceptionism for Targeted Molecule Generation. \textbf{2022}, \relax
\mciteBstWouldAddEndPunctfalse
\mciteSetBstMidEndSepPunct{\mcitedefaultmidpunct}
{}{\mcitedefaultseppunct}\relax
\EndOfBibitem
\bibitem[Bickerton \latin{et~al.}(2012)Bickerton, Paolini, Besnard, Muresan,
  and Hopkins]{bickerton2012quantifying}
Bickerton,~G.~R.; Paolini,~G.~V.; Besnard,~J.; Muresan,~S.; Hopkins,~A.~L.
  Quantifying the chemical beauty of drugs. \emph{Nature chemistry}
  \textbf{2012}, \emph{4}, 90--98\relax
\mciteBstWouldAddEndPuncttrue
\mciteSetBstMidEndSepPunct{\mcitedefaultmidpunct}
{\mcitedefaultendpunct}{\mcitedefaultseppunct}\relax
\EndOfBibitem
\bibitem[O'Boyle \latin{et~al.}(2011)O'Boyle, Banck, James, Morley,
  Vandermeersch, and Hutchison]{o2011open}
O'Boyle,~N.~M.; Banck,~M.; James,~C.~A.; Morley,~C.; Vandermeersch,~T.;
  Hutchison,~G.~R. Open Babel: An open chemical toolbox. \emph{Journal of
  cheminformatics} \textbf{2011}, \emph{3}, 1--14\relax
\mciteBstWouldAddEndPuncttrue
\mciteSetBstMidEndSepPunct{\mcitedefaultmidpunct}
{\mcitedefaultendpunct}{\mcitedefaultseppunct}\relax
\EndOfBibitem
\bibitem[Zhou \latin{et~al.}(2019)Zhou, Kearnes, Li, Zare, and
  Riley]{zhou2019optimization}
Zhou,~Z.; Kearnes,~S.; Li,~L.; Zare,~R.~N.; Riley,~P. Optimization of molecules
  via deep reinforcement learning. \emph{Scientific reports} \textbf{2019},
  \emph{9}, 1--10\relax
\mciteBstWouldAddEndPuncttrue
\mciteSetBstMidEndSepPunct{\mcitedefaultmidpunct}
{\mcitedefaultendpunct}{\mcitedefaultseppunct}\relax
\EndOfBibitem
\bibitem[Xie \latin{et~al.}(2021)Xie, Shi, Zhou, Yang, Zhang, Yu, and
  Li]{xie2021mars}
Xie,~Y.; Shi,~C.; Zhou,~H.; Yang,~Y.; Zhang,~W.; Yu,~Y.; Li,~L. MARS: Markov
  Molecular Sampling for Multi-objective Drug Discovery. International
  Conference on Learning Representations. 2021\relax
\mciteBstWouldAddEndPuncttrue
\mciteSetBstMidEndSepPunct{\mcitedefaultmidpunct}
{\mcitedefaultendpunct}{\mcitedefaultseppunct}\relax
\EndOfBibitem
\end{mcitethebibliography}
